\documentclass[12pt,a4paper]{scrartcl}
\usepackage[utf8]{inputenc}
\usepackage[T1]{fontenc}
\usepackage{lmodern}
\setkomafont{disposition}{\normalcolor\bfseries}
\usepackage{amsmath, amsfonts, amssymb, bbm, dsfont}
\usepackage{graphicx, color, enumitem, soul, verbatim}
\usepackage{geometry} 
\usepackage[sort]{natbib}
\usepackage{float, placeins, pdflscape}
\usepackage{multirow}
\usepackage{tabularx, booktabs, adjustbox}
\usepackage{threeparttable}
\usepackage{url}
\usepackage{rotating}
\usepackage{setspace}
\usepackage{subcaption}
\usepackage[titletoc,toc,title]{appendix}
\usepackage{booktabs}
\usepackage{multirow}

\usepackage[affil-it]{authblk}
\usepackage{hyperref}
\hypersetup{
colorlinks=true,
linkcolor=blue,
citecolor=blue
}
\newcommand{\red}[1]{\textcolor{red}{#1}}

\definecolor{mgray}{gray}{0.6}

\usepackage{tocloft}
   \cftsetindents{section}{3em}{3em}
   \cftsetindents{subsection}{4em}{4em}
   \cftsetindents{subsubsection}{5em}{5em}
   
\newcommand{\sym}[1]{#1} 

\usepackage[sectionbib]{chapterbib}      
\usepackage{bibunits}                    
\defaultbibliography{refs}       
\defaultbibliographystyle{chicagoa}       

\usepackage[en-US]{datetime2} 
\DTMlangsetup{showdayofmonth=false}

\usepackage{blindtext}
\usepackage{subfiles} 
   
\setlist{nosep}

\begin{document}

\begin{bibunit}
\maketitle
\thispagestyle{empty}

\begin{abstract}
Decomposing taxes by source (labor, capital, sales), we analyze the impact of 
automation on tax revenues and the structure of taxation in 19 EU countries during 1995-2016. 
Pre-2008, robot diffusion lead to decreasing factor and tax income, and a shift from taxes on capital to goods. ICTs changed the structure of taxation from capital to labor, with decreasing employment, but increasing wages and labor income. Post-2008, we find an ICT-induced increase in capital income and services, but no effect on taxation from ICT/robots. 
Overall, automation goes through various phases with heterogeneous economic effects which impact the amount and structure of taxes. Whether automation erodes taxation depends on the technology and stage of diffusion, and thus concerns about public budgets might be myopic when focusing on the short-run and ignoring relevant technological trends.



\end{abstract}

\vspace{3cm}
\noindent\textbf{\small Keywords:}{\small {} Automation, Technological Change, Taxation, Fiscal Revenues
\thispagestyle{empty} \setcounter{page}{0}\newpage{}\setcounter{page}{1} }{\small \par}


\newpage
\pagenumbering{arabic}
\setcounter{page}{1} 
\onehalfspacing

\section{Introduction}
\label{sec:introduction}
Does the diffusion of automation technologies (ATs) erode governments' tax basis? Taxes on labor contribute to a major share of public revenues. When ATs diffuse and replace labor at a large scale, the tax basis might be significantly undermined. This line of reasoning is put forward to argue that a tax on automation is needed to ensure the sustainability of public finances \citep{kovacev2020taxing, sussmuth2020taxation, acemoglu2020does, rebelo2019should}. 
However, the impact of automation is complex, including many second-order effects. In addition, governments receive taxes from multiple sources other than labor which might also be affected by ATs \cite[cf.][]{atkinson2019case}. Until now, there is limited empirical evidence on the nexus between automation and public revenues. This study aims to fill this gap. 
We explore empirical interactions between automation, production, and their link to taxation, to help understand the complex relationship between taxes and automation. 

Guided by a stylized model, we decompose tax revenues by source and link them to three 
effects of automation on economic activity called replacement, reinstatement and real income effects. First, the \emph{replacement effect} refers to all effects on factor demand and remuneration when human labor is replaced by sophisticated machinery able to execute tasks currently performed by humans \citep{brynjolfsson2014second, frey2017future, korinek2017artificial, acemoglu2020wrong, arntz2016risk, nedelkoska2018automation, gregory2018racing}. 
Second, the \emph{reinstatement effect} covers the creation of new tasks and occupations, and the reallocation of labor within and across industries \citep{acemoglu2019automation, blanas2019afraid, bessen2019automation, bessen2020firm, dauth2018adjusting}. Third, the \emph{real income effect} reflects changes in: (a) real income when reduced production costs affect prices; and (b) factor revenues, i.e. from capital and labor \citep{acemoglu2019automation, graetz2018robots, aghion2017artificial, korinek2017artificial}. 

The model serves as a conceptual framework to guide us through the interpretation of the empirical findings when addressing the following three research questions: 
\begin{enumerate}
		\item \emph{What is the relationship between AT diffusion and aggregate tax revenues at the country level?}
		\item \emph{What is the relationship between AT diffusion and the composition of taxes by source, i.e. labor, capital and goods?}
		\item \emph{How can these interactions be traced back to the three effects through which automation affects economic production?}
	\end{enumerate}

The complexity of tax systems and the multiple phases of technological change make it challenging to directly link the microeconomic impact of automation to macroeconomic consequences and aggregate taxation.
With this in mind, we rely on aggregate tax data from the \citet{OECD2020data} that allow for cross-country comparisons and to dissect tax accounts to taxes on labor, capital and goods for nineteen European countries during 1995-2016.
However, the effects of automation are expected to occur at the disaggregate level, i.e. firm and industry, when changes in production technology induce changes in factor demand, employees' incomes, and the level and composition of output. To understand these effects, we use macro- and industry-level production data from \cite{EUKLEMS2019data} for the same set of countries and periods considered in the tax data. 
To map technological and economic change at the micro-level to aggregate taxation, we base our analysis on country- and country-industry-level regressions. 
We start at the country-level by exploring interactions between automation and taxation along with the links between the structure of production and different tax sources. 
Next, we analyze the prevalence of the replacement, reinstatement, and real-income effect and argue how these effects help to explain the findings from above.

We find that the impact of automation depends on the technology type and the phase of diffusion. During the early phase (1995-2007), robots had a negative impact on aggregate tax revenues accompanied with decreasing factor income from capital and labor, and a shift from capital taxes to taxes on goods. 
After 2008, the effect of robots on factor markets and taxation disappears. 
Information and Communication Technologies (ICT) show different effects. Until 2007, we do not observe any effect on total revenues but an increasing demand for labor in automating industries. Country level employment effects are negative, but wages, capital prices, aggregate labor income, and output increased. We observe a change in the structure of taxation from capital to labor. After 2008, ICT diffusion shows a positive association with income and the employment share of the service sector.
However, we do not find any effect on total tax revenues. 

We conduct a battery of robustness checks to guard against various empirical concerns, such as: the late phase of AT diffusion coinciding with the 2008 global financial crisis; country-specific confounding factors, e.g. related to globalization and the tax system; and the potential endogeneity of the AT diffusion. 

Our results suggest that AT diffusion goes through different phases with different effects on taxes. Labor offsetting effects during an early phase may be compensated by the creation of new jobs in later periods accompanied with structural change in the industrial composition. 
However, given our observations, concerns about the sustainability of fiscal revenues appear short-sighted when only looking at the early phases of automation. 
Our framework provides structural arguments that enable a better understanding of the economic impacts of automation and macro-level effects on taxation. To the best of our knowledge, this is the first empirical study providing insights on the impact of automation on public finances.

The rest of the paper is structured as follows. In Section \ref{sec:background_taxation}, we provide an overview of the background on automation and taxation. 
In Section \ref{sec:model}, we introduce a conceptual model. In Section \ref{sec:empirical_approach}, we describe our empirical strategy and the data. Section \ref{sec:results} summarizes the results, while Section \ref{sec:robustness} provides a series of robustness checks. Section \ref{sec:discussion} discusses how the empirical results help to answer the research question and Section \ref{sec:conlusion} concludes.

\section{Background on taxation}
\label{sec:literature}

This section provides a description of tax systems in Europe, and an overview of the empirical and theoretical background on the link between taxation and automation. 

\subsection{Taxation in Europe}
\label{sec:background_taxation}
Taxes are {``compulsory, unrequited payments to general government''} \citep{OECD2019}. 
On average, among the nineteen European countries covered by our study, the total tax revenue accounted for 37.3\% of GDP in 2016 ranging from 23.4\% in Ireland to 45.7\% in Denmark.\footnote{When excluding residual taxes (with OECD-code $6000$), as done in our analysis, total taxes account for 37\% of GDP. Our analysis includes nineteen European countries: Austria (AT); Belgium (BE); Czech Republic (CZ); Germany (DE); Denmark (DK); Spain (ES); Finland (FI); France (FR); Greece (GR); Ireland (IE); Italy (IT); Lithuania (LT); Latvia (LV); the Netherlands (NL); Portugal (PT); Sweden(SE); Slovenia (SI); Slovakia (SK); and the United Kingdom (UK). The information presented here is based on the Global Revenue Statistics Database provided by the \citet{OECD2020data}.}
Over time, the average tax-to-GDP-ratio weakly fluctuated around 36.4\% in 1995 and 37\% in 2016, with the lowest ratio during the financial crises (e.g. 34.7\% in 2009). 

Taxes can be classified by the tax base. For example, taxes are imposed on income from labor, profits and capital gains, property, and trade of goods and services. Compulsory Social Security Contributions (SSC) can equally be considered as tax revenues \cite[][A.2]{OECD2019}. 
Here, we focus on three broad groups, namely taxes imposed on: (1) labor ($T^l$) including SSC; (2) capital ($T^k$) including taxes on profits and property; and (3) goods and services ($T^y)$. These groups differ by their linkage to the economic structure reflected in the labor share, capital share, and aggregate consumption. 

The three groups ($T = T^l + T^k + T^y$) cover more than 99.9\% of total tax revenue in our sample of nineteen European countries in 2016. 
On average, taxes on labor accounted for 11.8\% of GDP and 31.6\% of total taxation, taxes on capital for 13.3\% of GDP and 35.1\% of total taxation, and taxes on goods for 12\% of GDP and 32.5\% of total taxation.

Countries differ by the structure of taxation, i.e. the relative tax contribution of different sources. 
The cross-country heterogeneity in the levels, structure and organization of taxation is driven by a multitude of economic, structural, institutional, and social factors which have emerged historically across nations \citep{kiser2017political, hettich2005democratic, castro2014determinants}. Among others, empirical measures of such determinants include per-capita GDP, industrial structure and economic specialization, civil liberties and governmental efficiency, public and financial policies, trade, exchange rates, foreign direct investment, public expenditures and education \citep{castro2014determinants,castaneda2018tax}. We control for such relevant dimensions in our analysis.

\subsection{Taxation and automation}
For policy makers, two questions related to the nexus of automation and taxation are important: (1) How do existing tax systems influence AT adoption decisions and the emergent path of economic development?; and (2) Does automation affect tax revenues such that policy makers should care about their financial capacity
?
The majority of the existing literature on automation and taxation addresses the first question by taking as given that tax revenues suffice to finance essential public services. To the best of our knowledge, we are the first to study the second question. 

Existing studies on the nexus of automation and taxation mostly take an optimal taxation perspective. 
\cite{acemoglu2020does} argue that the US tax system is biased in favor of capital, which leads to a sub-optimal reduction of the labor share for ``marginally automated jobs''. Applying the optimal taxation framework by \cite{diamond1971optimal} to a task-based model calibrated on US tax rates, the authors show how a tax reform could raise the labor share. 
Similarly, \cite{sussmuth2020taxation} analyze the impact of US taxation on the functional distribution of income and find that distributional changes (in favor of the capital share) can be partly attributed to labor and capital tax reforms during 1974-2008. They argue that changes in relative taxes also affect the use of robots. 

Other authors propose a robot tax to cope with the negative effects of automation on employment and income equality. In a theoretical study based on the current tax system in the US, \cite{rebelo2019should} show how a robot tax can be used to reduce inequality, but at the cost of efficiency losses. \cite{gasteiger2017note} provide a theoretical analysis of a robot tax in overlapping generations model and show how it could raise per capita capital stock with positive long-run growth effects. 
\cite{kovacev2020taxing} argues theoretically that the robot-induced replacement of labor could lead to decreasing income taxes and higher transfer payments. A robot tax is analyzed as an instrument to offset the negative effect on public finances. From a law-and-economics perspective, he shows how a robot tax could be implemented while keeping the disincentives for innovation minimal.

Theoretical studies on robot taxes argue that these taxes can be used to reduce inequality and to secure public revenues. 
However, it remains controversial whether automation really undermines governments' capacity to raise taxes. \citet{atkinson2019case} argues that empirical evidence of a jobless future is poor since many studies ignore important second-order effects. Moreover, even if firms adopt robots they still pay taxes on profits, sales and wages of workers doing non-automated jobs. 

Up to date, empirical evidence on the relationship between automation and tax revenues is lacking and we aim to fill this gap. While studies on optimal taxation focus on the impact of tax systems on the structure of production, we take the opposite perspective and look at the impact of economic change on taxation. 
In contrast to optimal taxation studies, we do not look at relative tax rates, but study aggregate tax revenues. 
While changes in relative tax rates on labor and capital might have affected the diffusion of ATs in the US, as argued by \citet{acemoglu2020does}, we do not find evidence that changes in the tax system have been a driver of excessive automation in Europe. Specifically, using data on implicit tax rates on labor and capital, we find that these rates remained roughly constant in most European countries during the past decade.\footnote{See Appendix Figure \ref{fig:rel_taxes_c}. The data on implicit tax rates on labor and capital in Europe, provided by the European Commission, are not directly comparable to the approach used by \citet{acemoglu2020does} who calculate effective tax rates on labor and different types of automation technologies at the micro-level in the US. Note that it is not straightforward to apply their methodology in a European cross-country setting with very heterogeneous and complex tax systems. The European Commission computes implicit tax rates on labor and capital as a ratio of actual tax income by source to the potential tax basis \citep{EC2020taxreport}. Nonetheless, the stable patterns on relative tax (rates) observed in the data for the EU are in stark contrast to the clear-cut divergence in favor of capital observed by \citet{acemoglu2020does} for the US.}  
Moreover, our results suggest different diffusion patterns for robots and ICT (see Figure \ref{fig:key_vars_all_regions}) indicating that the effects we study are unrelated to relative tax rates.

\section{A conceptual framework}
\label{sec:model}
This section provide a stylized model to decompose tax revenues by source and link them to the three interdependent effects of automation on economic activity: replacement and reinstatement of labor and changes in real income.

\subsection{Tax revenues}
Taxes can be grouped into three parts that are differently linked to production. Total tax revenue in country $c$ is given by: 
\begin{align}
T_c = \underbrace{t^l_c \cdot w_c L_c}_{\begin{subarray}{l}\text{Taxes on labor}\\ 
\text{\hspace{0.75cm}$T^l_c$}\end{subarray}} + \underbrace{t^k_c \cdot r_{c} K_c}_{\begin{subarray}{l}\text{Taxes on capital}\\ 
\text{\hspace{0.75cm}$T^k_c$}\end{subarray}} + \underbrace{t^{y}_c \cdot p_c Q_c}_{\begin{subarray}{l}\text{Taxes on goods}\\ 
\text{\hspace{0.75cm}$T^y_c$}\end{subarray}}
\label{equation:tax_revenues}
\end{align}
where $L_c = \sum_{i \in I_c} L_{i,c}$ is aggregate labor given by the sum of labor employed in industries $i \in I_c$ in country $c$, $K_c = \sum_{i \in I_c} K_{i,c}$ is the fixed production capital stock including ATs (industrial robots and ICT), and $p_c Q_c = \sum_{i \in I_c} p_{i,c}Q_{i,c}$ is aggregate demand. Wages, capital prices and goods prices are given by $w_c$, $r_{c}$ and $p_c$, respectively. 
The tax rates $t^{l}_c$, $t^{k}_c$ and $t^{y}_c$ are imposed on labor income, capital income and final demand, respectively.

\subsection{Production technology}
Automation changes firms' production technology. This can have an impact on industry-level factor demand, i.e. labor and capital, and productivity when firm-specific production processes and industrial organization change. 
In a generic form, the production function of industry $i$ is:
\begin{align}
Y_{i,c} = f_{i,c}(K_{i,c}, L_{i,c}, A_{i,c}) 
\end{align}
with $K_{i,c}$ and $L_{i,c}$ as the respective capital and labor whose demand depends on wages $w_{i,c}$ and capital prices $r_{i,c}$, respectively. The capital stock $K_{i,c}$ comprises of different types of capital, i.e. $K_{i,c} = K^n_{i,c} + K^a_{i,c}$ where $K^n_{i,c}$ is non-automation capital and $K^a_{i,c} =  ICT_{i,c} + R_{i,c}$ is automation capital with $R_{i,c}$ as industrial robots and $ICT_{i,c}$ as ICTs.\footnote{$ICT_{i,c}$ and $R_{i,c}$ are not necessarily disjoint.} Both, robots and ICT, are measures of automation, but capture different concepts. Industrial robots are pure ATs designed to automate manual tasks performed by human workers. ICT capital is more general and can be used for various cognitive tasks complementing or substituting human labor. 
We assume that all types of capital are rented at the same price $r_{i,c}$. 

Production technologies differ across industries and countries leading to different input shares. 
Production functions are empirically not observable. However, we observe industry-level factor inputs, factor costs and output, which allow us to draw inference about the relationships between inputs, outputs and the price responsiveness of factor demand. 
By the definition of a production function, we assume $\tfrac{\partial f_{i,c}}{\partial L_{i,c}} \geq 0$, $\tfrac{\partial f_{i,c}}{\partial K_{i,c}}\geq 0$, $\tfrac{\partial f_{i,c}}{\partial A_{i,c}} \geq 0$, i.e. the quantity of output is non-decreasing in the quantity of inputs and in the level of productivity. 
Moreover, ceteris paribus, we expect factor demand to be negatively related to factor prices, i.e. $\tfrac{\partial L_{i,c}}{\partial w_{i,c}} \leq 0$ and $\tfrac{\partial K_{i,c}}{\partial r_{i,c}} \leq 0$.

\subsection{Final demand}
Final demand is given by the aggregation across firms: 
\begin{align}
    p_c Q_c = \sum_{i \in I_c} p^s_{i,c} q_{i,c}(p_{i,c}, Y_c)
\end{align}
where $p_{i,c} = (1+t^y)* p^s_{i,c}$ is $i$'s consumer price including consumption taxes $t^y$, $p^s_{i,c}$ is $i$'s supply price, and $q_{i,c}(p_{i,c}, Y_c)$ is industry level demand which is a function of the price and income $Y_c$ in region $c$ with $\tfrac{\partial q_{i,c}}{\partial p_{i,c}}\leq 0$ and $\tfrac{\partial q_{i,c}}{\partial Y_c}\geq0$. Assuming market closure, income is composed of labor income $w_c L_c$, capital income $r_c K_c$ minus tax payments, such that:
\begin{align}
     Y_c = (1-t^l_c) \cdot w_c L_c + (1 - t^k) \cdot r_c K_c.
\end{align}

In this stylized representation, we abstain from trade, inter-regional transfers, savings, and household and firm heterogeneity.

\subsection{Effects of automation}
Automation indirectly affects tax revenues through changes in production technology that translate into changes in factor use, market shares, and final demand.

Formally, the aggregate effect on tax revenue is given by the differential 
\begin{align} \nonumber
d T_c = t^l_c \cdot &\left( \frac{\partial w_c}{\partial K^a_c} L_c + w_c \frac{\partial L_c}{\partial K^a_c} \right) + t^k_c \cdot \left(\frac{\partial r_c}{\partial K^a_c} K_c + r_c \frac{\partial K_c}{\partial K^a_c} \right) 
 \\[0.25cm] &+ t^{Y} \cdot \left(\frac{\partial P_c}{\partial K^a_c} Q_c + P_c \frac{\partial Q_c}{\partial K^a_c}\right)
 \label{equation:differential_tax_revenues}
\end{align}
where $K^a_c = R_c + ICT_c$, with $R_c = \sum_{i \in I_c} R_{i,c}$ and $ICT_c = \sum_{i \in I_c} ICT_{i,c}$.

We study the effect of automation on production and taxation along three effects: replacement, reinstatement, and real-income. 
Even if the distinction between these effects is not clear-cut, we simplify the analysis and assume the replacement and reinstatement effect to be mainly reflected in a changing factor demand, while the real income effect to be reflected in final demand and prices. We next discuss these effects in detail.

\subsubsection{Replacement} 
The replacement effect is the substitution of human labor by machines when technological progress enables machines to perform tasks previously performed by humans \citep{acemoglu2018artificial}. 
The number of jobs susceptible to automation differs across occupations and industries \citep{arntz2016risk, frey2017future, nedelkoska2018automation, pwc2018, webb2019impact}. 
Labor replacement may lead to lower employment and wages which may be offset by an increase in demand for non-routine tasks and new jobs in expanding sectors \citep{acemoglu2019automation, acemoglu2020wrong, acemoglu2018modeling, acemoglu2018artificial}. 

Empirical results on the labor replacement effects remain ambiguous \citep{gregory2018racing, graetz2018robots, aghion2019innovation, blanas2019afraid, dauth2018adjusting}. 
Overall, it is consensual in the literature that employees performing automatable tasks are susceptible to replacement by machinery, but it remains controversial whether and to what extent the occupation-specific replacement affects aggregate factor incomes. 

In automating industries, characterized by $K^a_{i,c} > 0$, employees are potentially replaced by machinery with $\tfrac{\partial L_{i,c}}{\partial K^a_{i,c}} < 0$ for $i \in \{ j | K^a_{j,c} > 0 \}$. 
The effect on wages in industry $i$ can go either way: $\tfrac{\partial w_{i,c}}{\partial K^a_{i,c}} \lessgtr 0$: 
On the one hand, the replacement effect exerts downward pressure on wages paid for jobs that can be automated. On the other, automation may complement non-automatable labor which increases productivity with a positive effect on wages and leads to a polarization of wage income. 
The net impact of the replacement effect on the labor income in industry $i$ depends on the extent to which potential wage increases for non-automatable jobs offset the replacement of automatable jobs, whereby we expect a negative sign if the replacement dominates reinstatement in industry $i$ giving $\tfrac{\partial (w_{i,c}L_{i,c})}{\partial K^a_{i,c}} < 0$.

Ceteris paribus, in the absence of the reinstatement and real-income effect, the replacement effect has a negative impact on total and labor taxes in particular, provided that the net effect on the wage bill is negative and taxes are sufficiently non-progressive. If labor taxes are progressive, taxes on labor and wage polarization are positively related. 

\subsubsection{Reinstatement}
Historically, job replacement by automation was often compensated by the emergence of new occupations and the reinstatement of labor \citep{acemoglu2019automation, autor2015there, aghion2017artificial, bessen2019automation}. 
Reinstatement effects occur at different levels of analysis. Within automating industries, automation may induce occupational changes driven by two effects: (1) efficiency gains may release resources available for other processes, and (2) automation may require complementary labor inputs to operate the machinery. 
This effect can be reinforced if automation stimulates capital accumulation which may also have a positive effect on labor demand. 

The reinstatement effect can also occur as a spillover at the aggregate level when productivity growth reduces prices or increases in incomes lead to market growth and/or changing market shares and sizes of other industries. This can induce the reinstatement of labor in other industries and a cross-industrial reallocation of labor. 
The employment and income effects may differ across industries, skill, and occupational groups, and the process of reinstatement may be slowed down by labor market frictions and skill mismatches \citep{acemoglu2020wrong,  bessen2020firm, gregory2018racing, dauth2018adjusting, arntz2016risk}. 

The reinstatement effect potentially offsets sector-specific negative employment effects at the aggregate level. 
Ceteris paribus, the reinstatement effect positively affects labor demand in automating industries and at the country level, i.e. $\tfrac{\partial L_{i,c}}{\partial K^a_{i,c}}>0,  i \in \{ j | K^a_{j,c} > 0 \}$ and $\tfrac{\partial L_c}{\partial K^a_c}>0$. 
Dependent on wage heterogeneity within and across industries, the reinstatement effect can have ambiguous effects on industry and country level average wages. However, it has a positive effect on aggregate labor income, i.e. $\tfrac{\partial w_cL_c}{\partial K^a_c} > 0$. 

\subsubsection{Real income}
The real income effect is an indirect, composite effect resulting from the replacement and reinstatement of labor, and the impact of automation on capital accumulation, productivity and prices. 
Automation may boost productivity growth. This can lead to lower output prices and leverage economic growth through increasing final demand \citep{graetz2018robots, acemoglu2018artificial, gregory2018racing}. Final demand is contingent on real income, i.e. nominal income over prices \citep{bessen2019automation}. Both may be affected by automation.  
Whether productivity-induced cost reductions are transmitted to consumers as lower prices is contingent on market competition which might be undermined by an unequal distribution of the benefits of AT diffusion \citep{autor2020fall, bormans2020link, andrews2015frontier, andrews2016global, barkai2020declining}.

The direction of the total effect of automation on aggregate nominal income depends on the relative contribution of the different mechanisms, $\tfrac{\partial (w_cL_c + r_cK_c)}{\partial K^a_c} \lessgtr 0$. 

The second part of the real income effect is a productivity induced change in the aggregate price level. Productivity has a negative effect on unit production costs. ATs increase productivity, i.e. $\tfrac{\partial A_{i,c}}{\partial K^a_{i,c}} \geq 0$ assuming rational AT adoption decisions, which leads to price reductions when lower unit production costs are passed through to consumers, i.e. $\tfrac{\partial p_{i,c}}{\partial A_{i,c}} \leq 0$ and $\tfrac{\partial p_{i,c}}{\partial K^a_{i,c}} \leq 0$. In turn, this increases real disposable income, i.e. $\tfrac{\partial Y^r_c}{\partial K^a_{i,c}} \geq 0$ where $Y^r_c=(1-t^l) \tfrac{w_c}{p_c}L_c + (1-t^k) \tfrac{r_c}{p_c}K_c$ and $\tfrac{\partial p_c}{\partial p_{i,c}} \geq 0$ and $\tfrac{\partial p_{i,c}}{\partial K^a_{i,c}} \leq 0$.

Dependent on the income elasticity of demand, an increase in real income may induce an increase in consumption which reinforces the reinstatement effect with positive feedback on labor and capital income.

\FloatBarrier
\section{Empirical approach and data}
\label{sec:empirical_approach}
In this section, we provide an overview of the steps followed to empirically examine the effects of interest, and in turn discuss the relevant variables and data sets used.

\subsection{Overview}
Real-world tax systems are complex. 
Tax revenues are raised through different channels with many non-linearities arising from diverse threshold levels and exemptions. Uniform and linear macroeconomic tax rates $t^l_c$, $t^k_c$ and $t^y_c$ as suggested by our theoretical framework do not exist. While, data on taxation is only available at the country level, tax burdens are heterogeneous across households, firms, and industries. However, many of the effects of automation occur at the industry or firm level. Therefore, to analyze the effect of automation on taxation, we use an indirect approach. 
Empirically, we observe tax revenues (i.e. $T$, $T^l$, $T^k$, $T^y$) at the country level, measures for key economic variables (i.e. $w$, $L$, $r$, $K$, $p$, $Q$) at the country and country-industry level, and indicators for the economic structure across periods $t$.

Our procedure consists of the following steps. 
First, we establish prerequisites that motivate the subsequent steps. This includes testing for  associations between taxes and automation and examining the empirical link between different types of taxes and economic variables.
Second, explore the prevalence of each of the three effects: replacement, reinstatement, and real income; summarized in Table \ref{tab:automation_effects}.
Finally, we argue how the three effects help to explain the impact of automation on taxation and in turn help us answer the three research questions introduced in Section \ref{sec:introduction}. 

\begin{table}[H] \centering
\caption{Overview of three key effects of automation on economic production}
\label{tab:automation_effects}
\footnotesize
\begin{tabular}{p{2cm}p{7.4cm}p{4cm}}
\toprule
     Effect&Description&Indicators  \\
\midrule
     Replacement&Substitution of labor. Decreasing labor demand and wages. Unclear side effects on net capital accumulation, prices and depreciation.& $\tfrac{\partial L_{i,c}}{\partial K^a_{i,c}}$, $\tfrac{\partial w_{i,c}}{\partial K^a_{i,c}}$, $\tfrac{\partial r_{i,c}}{\partial K^a_{i,c}}$, $\tfrac{\partial K_{i,c}}{\partial K^a_{i,c}}$ where $K^a_{i,c} = R_{i,c} + ICT_{i,c}$ and $i \in \{j | K^a_{j,c} > 0\}$.
     \\
\midrule
    Reinstatement&Productivity gains from automation reinstate labor demand in other/newly emerging economic activities. Increasing labor demand and wages.&$\tfrac{\partial L_c}{\partial K^a_c}$, $\tfrac{\partial w_c}{\partial K^a_c}$, $\tfrac{\partial r_c}{\partial K^a_c}$, $\tfrac{\partial K_c}{\partial K^a_c}$, $\tfrac{\partial Services_c}{\partial K^a_c}$.
    \\
\midrule 
    Real income&Productivity gains reduce unit production costs and prices of final goods, and increase aggregate demand. Distortions in market structure and competition, and an unequal distribution of income gains may undermine this effect.&$\tfrac{\partial A_c}{\partial K^a_c}$, $\tfrac{\partial p_c}{\partial K^a_c}$, $\tfrac{\partial Q_c}{\partial K^a_c}$, $\tfrac{\partial HHI_c}{\partial K^a_c}$.\\
\bottomrule
\end{tabular}
\end{table}

\subsection{Data}
We combine different data sets at different aggregation levels with varying coverage in terms of countries, industries and time. After merging the data
, we end up with two samples covering nineteen European countries for the period 1995-2016.\footnote{List of countries: AT; BE; CZ; DE; DK; ES; FI; FR; GR; IE; IT; LT; LV; NL; PT; SE; SI; SK; UK.} 
The first sample is a country-level panel for the whole economy, i.e. from agriculture to public sectors. 
The second sample is an industry-level panel covering only automation-intensive industries, i.e. industries where information on the use of robots exists by the International Federation of Robotics (IFR) \citep{IFR2020data}. The automation-intensive industries include: agriculture; mining and quarrying; ten manufacturing industries; electricity, gas and water supply; construction; and education, research and development (see Appendix Table \ref{tab:industry_codes}.)

\begin{figure}[H]
{\centering
\caption{Time series of key variables (averaged across countries)}
\label{fig:key_vars_all_regions}
\includegraphics[width=1\textwidth]{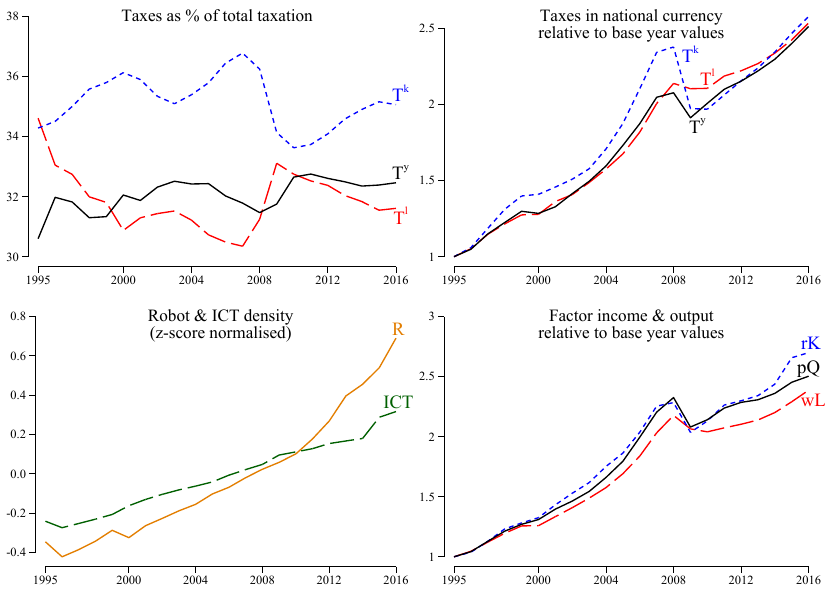}
}
\scriptsize
Source: Author's calculations based on IFR, EUKLEMS and OECD Global Revenue Statistics Database.\\
Notes: Each time series represents the average value of the respective variable across all 19 European countries considered in the country-level sample. $T_{c,t}^l$, $T_{c,t}^k$ and $T_{c,t}^y$ refer to taxes on labor, capital and goods, respectively. $R$ and $ICT$ capture the robot and ICT intensity as the ratio of the number of operational robots and ICT capital, respectively, over the number of hours worked in the economy. $wL$, $rK$ and $pQ$ is labor compensation, capital compensation and the value of gross output, respectively. For the two right panels, 
the country-level values of each variable considered are indexed relative to their base year values. For the bottom-left panel, $R$ and $ICT$ are z-score normalized by subtracting the sample mean and dividing by the standard deviation of the sample. The sample includes nineteen European countries: AT; BE; CZ; DE; DK; ES; FI; FR; GR; IE; IT; LT; LV; NL; PT; SE; SI; SK; and UK, for the period 1995-2016, but is unbalanced since data are not reported for LT, LV and UK in 1995, and DK, PT, SI and SK in 1995-1999. For more details over the country-level sample and construction of variables, see Online Appendix Section~\ref{sec:empirical_approach_tax}.
\end{figure}

\subsubsection{Tax revenue}
\label{sec:empirical_approach_tax}
Taxes are part of our country-level data compiled from the OECD Global Revenue Statistics Database \citep{OECD2020data}. We retrieve information on taxes by type, i.e. labor ($T_{c,t}^l$), capital ($T_{c,t}^k$), and goods ($T_{c,t}^y$), measured in national currency and as percentage of GDP or total taxation.
Time series plots are shown in the top panels of Figure \ref{fig:key_vars_all_regions}. A kink during the 2008 financial crisis is visible in both relative tax contributions from different sources and total tax revenues. The financial crisis was associated with a significant decline in capital tax revenues leading to a relatively larger relative tax burden on labor and goods.

\subsubsection{Economic variables}
\label{sec:empirical_approach_economic_vars}
Empirical proxies for factor income and consumption at the country level are aggregates of NACE Rev. 2 (ISIC Rev. 4) industry-level data from the EUKLEMS database \citep{EUKLEMS2019data, stehrer2019industry, adarov2019tangible}. 
The bottom right panel in Figure \ref{fig:key_vars_all_regions} illustrates the evolution of the macroeconomic accounts $wL_t$, $rK_t$, $pQ_t$ averaged across all countries and normalized to the base year 1995. We see that aggregate revenues from labor increased slower compared to the other accounts.\footnote{Along with other related sources, we construct a host of indicators used to ensure the robustness of our analysis. For details over the construction and use of variables, see Online Appendix \ref{appendix:data}.}

\subsubsection{Measuring automation technologies}
\label{sec:empirical_approach_ai_data}
We rely on two measures of automation based on: (1) the number of operational industrial robots from \citet{IFR2020data}; and (2) the capital stock of ICT from \citet{EUKLEMS2019data}. 
To capture the extent to which robots were incorporated in production technologies, we closely follow \cite{graetz2018robots} to construct the \emph{robot density} measure as the number of operational robots over the number of hours worked by human labor. 
Similarly, as a second automation indicator, we use the \emph{ICT-intensity} measured as net ICT capital stock per hour worked. These measures are computed both at the country-year and industry-year dimension. 


We consider these two measures to account for two distinct AT types distinguished by the type of task they execute. Specifically, robots are designed to perform manual tasks, while ICTs have a stronger link to cognitive tasks. 
While robots are pure ATs that execute a well defined task previously performed by humans, it is less clear whether this also applies to ICTs. ICTs can be flexibly applied in many different tasks and, to some extent, these tasks do not necessarily have a clear analogue in the range of tasks executed by humans. 

In our analysis, we use both measures simultaneously and as an interaction term. Robot-ICT interaction, otherwise stated as the \emph{depth of automation}, captures complementarities between the two ATs, i.e. the extent to which both manual and cognitive tasks are performed by machinery. Concerns about multicollinearity are ruled out since the correlation between measures is low (i.e. correlation coefficient of 0.22). 
The bottom left panel in Figure \ref{fig:key_vars_all_regions} presents a time series plot of the z-score normalized measure of robot- and ICT-intensity and suggests that post-2008 the rate of robot diffusion outpaced that for ICTs which exerts a stable rise since 1995.


\section{Results}
\label{sec:results}

\subsection{Taxation, automation and the economy}
\label{sec:results_prerequisites}
Before analyzing the channels through which automation affects the economy, we describe empirically the nexus between taxation, macroeconomic aggregates and automation. 

We begin by regressing country level tax revenues on AT diffusion measures and key indicators that describe the structure of economic production, i.e. 
\begin{equation}
    \mathbb{T}_{c,t} = \beta^R R_{c,t} + \beta^{ICT} ICT_{c,t} + \beta^{RICT} R_{c,t} * ICT_{c,t} + \beta^L wL_{c,t} + \beta^K rK_{c,t} + \beta^z  Z_{c,t} + \epsilon_{c,t} \label{eq:taxrevenue}
\end{equation}
where $\mathbb{T}_{c,t} \in \{ T_{c,t}, T^l_{c,t}, T^k_{c,t}, T^y_{c,t} \}$ reflects taxes in (1) levels, i.e. logs of billions of national currency, (2) percentage share of GDP and (3) percentage share of total taxation. We control for country and time FE and a set of macroeconomic variables $Z_{c,t}$ to account for country-specific economic characteristics, global shocks, and various potential confounding factors that could be driving taxes and are correlated with the AT diffusion measures, respectively.\footnote{These include: GDP growth; gross output share of service industries; Herfindahl-Hirschman Index based on the gross output shares of macro-sectors; government consolidated gross debt as \% of GDP; government interest payable as \% of GDP; net government lending/borrowing as \% of GDP; gross fixed capital formation as \% of GDP; and period average exchange rate. All regressions for Taxes in ln of national currency also include the ln of gross output value ($pQ$). For more details over the construction and use of these variables, see Online Appendix~\ref{appendix:data}.} To allow the error term to be correlated both across countries and over time, we cluster standard errors at the country and time dimension.

Results are presented in Table \ref{tab:regr_pre_tax_robots} for various time periods (Panels A-C).\footnote{For space considerations, we only present the variables of interest in the main text, but provide and discuss results about the remaining determinants in the Online Appendix Section \ref{appendix:add_results_determinants_taxation}.} 
In the first block of columns, we see the association of automation with taxes measured in logarithmic national currency units. The second block shows the relationship with taxes measured in percentage GDP. The last block shows the impact on the structure of taxation, i.e. on taxes as a share of total taxation. Labor and capital income ($wL_{c,t}$ and $rK_{c,t}$) are measured in levels in the first block and as a percentage share in total output $pQ_{c,t}$ in the latter two blocks to proxy for the labor and capital share. 

For the full period (Panel A), we observe a negative relationship between robots and total tax revenues and taxes on labor in absolute terms. Until 2007 (Panel B), robot diffusion was associated with a decline in capital taxation while increasing the share of taxes on goods. After 2008 (Panel C), we do not find any significant association between robots and taxes. 
For ICT, we observe a negative relationship with total tax revenues. Here, the decline in taxes comes hand in hand with reductions in capital taxes while we observe an increasing share of taxes on goods. 
The depth of automation, captured by the interaction term, exhibits a positive relationship with total tax revenues and taxes on capital. This is also reflected in the structure of taxation where we observe a shift from taxes on labor to capital.

\begin{table}[H]
\caption{Taxation and automation}
{\centering
\begin{adjustbox}{width=\textwidth}\begin{tabular}{l cccc|cccc|ccc}
\toprule \addlinespace
\csname @@input\endcsname{inputs/all_95_16/Results2shortV1_RICTint_ct_All_95t16}
\midrule
\csname @@input\endcsname{inputs/all_subperiod/Results2shortV1_RICTint_ct_All_95t07}
\midrule
\csname @@input\endcsname{inputs/all_subperiod/Results2shortV1_RICTint_ct_All_08t16}
\bottomrule
\end{tabular}\end{adjustbox}} \\ 
\label{tab:regr_pre_tax_robots} \scriptsize
Notes: \sym{*}\(p<0.05\), \sym{**}\(p<0.01\), \sym{***}\(p<0.001\). 
All regressions use country-level data for nineteen European countries during 1995-2016 and include: GDP growth, gross output share of service industries; Herfindahl-Hirschman Index based on the gross output shares of macro-sectors; government consolidated gross debt as \% of GDP; government interest payable as \% of GDP; net government lending/borrowing as \% of GDP; gross fixed capital formation as \% of GDP; period average exchange rate; and country ($c$) and year ($t$) fixed effects. Regressions for Taxes in ln of national currency also include the ln of gross output value ($pQ$). Standard errors are clustered both at the country and year level.
\end{table}

Also, we include labor and capital income (shares) as additional controls. Based on theory and empirical findings from existing studies, we expect automation to affect taxation through the channels of production, income and distribution. 
We observe a strong positive relationship between the wage bill and all taxes measured in national currencies. Looking at taxes measured in \% of GDP and the structure of taxation, we find an increasing labor share to be negatively associated with total taxes and taxes on goods. 
This pattern is robust and holds across different time periods. Capital income exhibits similar associations, but appears to be unrelated to taxes in levels. 


\FloatBarrier
\subsection{The impact of automation on the economy}
\label{sec:results_ai_effects}
\FloatBarrier
\subsubsection{Replacement effect}
\label{sec:results_replacement}
\label{sec:empirical_approach_replacement}
We test for the replacement effect running the following regressions:
\begin{equation}
     \mathbb{X}_{i,c,t} = \beta^R R_{i,c,t} + \beta^{ICT} ICT_{i,c,t} + \beta^{RICT} R_{i,c,t} * ICT_{i,c,t} 
    + \epsilon_{i,c,t} 
\end{equation}
where $ \mathbb{X}_{i,c,t} \in \{wL_{i,c,t}, w_{i,c,t}, L_{i,c,t}, rK_{i,c,t}, r_{i,c,t}, K_{i,c,t} \}$ refer to the values, prices and quantities of labor and capital, and $i$ refers to an automation-intensive industry (see \ref{sec:empirical_approach_ai_data}). 
We control for country-industry, country-year and industry-year FE to account for unobserved heterogeneity across those dimensions and look at changes over time within country-industries. Standard errors are two-way clustered at the country-industry and year level.

Table \ref{tab:regr_replacement} presents results on the replacement effect.
In total, we find weak support for the replacement effect in automation-intensive industries when robots diffuse, i.e. we observe decreasing employment aligned with increasing wages. The effects offset such that we do not find a significant effect on the wage bill $wL_{i,c,t}$. The replacement effect is stronger when robots and ICT are adopted simultaneously as captured by $R \ast ICT_{i,c,t}$, as there is an additional negative effect on top of the robot-specific one. Nonetheless, the net effect on total labor is weak. 
ICT diffusion is weakly positively associated with $wL_{i,c,t}$ driven by increasing employment $L_{i,c,t}$. This association is only significant prior to 2007. 
After 2008, we do not observe strongly significant effects on labor market outcomes in automation-intensive industries. 
Also, we do not find that automation has any significant association with capital accumulation and valuation. 

\begin{table}[H]
{\centering 
\caption{Replacement effect}
\label{tab:regr_replacement}
\begin{adjustbox}{width=0.6\textwidth}\begin{tabular}{l ccc ccc}
\toprule \addlinespace

\csname @@input\endcsname{inputs/all_95_16/Results3RICTint_cit_All_95t16}
\bottomrule \addlinespace
\csname @@input\endcsname{inputs/all_subperiod/Results3RICTint_cit_All_95t07}
\bottomrule \addlinespace
\csname @@input\endcsname{inputs/all_subperiod/Results3RICTint_cit_All_08t16}
\bottomrule
\end{tabular}\end{adjustbox}\\ }
\scriptsize
Notes: \sym{*}\(p<0.05\), \sym{**}\(p<0.01\), \sym{***}\(p<0.001\). 
All regressions use industry-level data for nineteen European countries during 1995-2016 for the set of industries susceptible to automation, 
and include: country-industry ($ci$); country-year ($ct$); and industry-year ($it$) fixed effects. All regressions are weighted by the base-sample-year share of each industry’s number of hours worked to country-wide hours worked. Standard errors are clustered both at the country-industry and year level.
\end{table}

\FloatBarrier
\subsubsection{Reinstatement effect}
\label{sec:results_reinstatement}

We empirically test the reinstatement effect with the following country level regressions:
\begin{equation}
    \mathbb{Y}_{c,t} = \beta^R R_{c,t} + \beta^{ICT} ICT_{c,t} + \beta^{RICT} R_{c,t} * ICT_{c,t}+ \beta^z  Z_{c,t} + \epsilon_{c,t} 
\end{equation}
where $\mathbb{Y}_{c,t} \in \{ w_{c,t}, L_{c,t}, r_{c,t}, K_{c,t}, Services_{c,t}, Gini^w_{c,t} \}$. 
The main effects of interest are those of automation on labor market outcomes $w_{c,t}$ and $L_{c,t}$. 
Moreover, we examine qualitative features of the effect by testing whether automation is a driver of capital accumulation ($K_{c,t}$ and $r_{c,t}$) and the cross-industrial reallocation of output from goods to services captured by the output share of services $Services_{c,t}$. With $Gini^w_{c,t}$ we evaluate potential effects on cross-industrial wage inequality captured by the Gini coefficient. The regressions include the same country-level controls $Z_{c,t}$ as in equation~\eqref{eq:taxrevenue}, and country and year FE. We cluster standard errors at the country and year level. Regression results are shown in Table \ref{tab:regr_reinstatement}.

\begin{table}[H]
{\centering 
\caption{Reinstatement effect}
\label{tab:regr_reinstatement}
\begin{adjustbox}{width=0.6\textwidth}\begin{tabular}{l ccccccc}
\toprule \addlinespace
\csname @@input\endcsname{inputs/all_95_16/Results4shortRICTint_ct_All_95t16}
\bottomrule \addlinespace
\csname @@input\endcsname{inputs/all_subperiod/Results4shortRICTint_ct_All_95t07}
\bottomrule \addlinespace
\csname @@input\endcsname{inputs/all_subperiod/Results4shortRICTint_ct_All_08t16}
\bottomrule
\end{tabular}\end{adjustbox}\\ }
\scriptsize
Notes: \sym{*}\(p<0.05\), \sym{**}\(p<0.01\), \sym{***}\(p<0.001\). 
All regressions use data for nineteen European countries during 1995-2016 and include: GDP growth, government consolidated gross debt as \% of GDP; government interest payable as \% of GDP; net government lending/borrowing as \% of GDP; gross fixed capital formation as \% of GDP; period average exchange rate; value added TFP--calculated as the residual from an OLS regression of value-added volumes ($VA$) on a translog production function including capital volumes ($K$) and total number of hours worked ($L$); and country ($c$) and year ($t$) fixed effects. Standard errors are clustered both at the country and year level.
\end{table}

We observe that robots and the depth of automation exhibit a negative association with wages and capital prices. These effects are most significant prior to 2007. 
Robots alone are negatively associated with the demand for capital (and insignificantly with labor). In contrast, we find a positive association of the automation depth with the demand for labor and (insignificantly) for capital. 
As above, the effects are only significant prior to 2007 except from the negative association between automation depth and wages.

For ICT diffusion, we observe different patterns. ICT is significantly positively associated with factor prices, but negatively with factor demand quantities, for both labor and capital. After 2008, the relationships between ICT diffusion and factor demand quantities become positive, but are only weakly significant for capital.  

Generally, we observe that the diffusion of automation technologies is negatively associated with the output share of services.
This effect is significant for robots in both periods. 
ICT exhibits a significant negative correlation with the service share prior to 2007, but we do not see any significant effect after 2008. 
For the whole period, we find that both automation technologies exhibit a positive correlation with cross-industrial wage inequality.

\FloatBarrier
\subsubsection{Real income effect}
\label{sec:results_real_income}
\label{sec:empirical_approach_real_income}
To evaluate the real income effect of automation on: (1) aggregate factor incomes; and (2) productivity and prices of final goods $p_{c,t}$, while accounting for aggregate market expansion reflected in aggregate output $Q_{c,t}$ and sales $pQ_{c,t}$, we run the following regressions: 
\begin{equation}
    \mathbb{Y}_{c,t} = \beta^R R_{c,t} + \beta^{ICT} ICT_{c,t} + \beta^{RICT} R_{c,t} * ICT_{c,t} + \beta^z  Z_{c,t} + \epsilon_{c,t} 
\end{equation}
where $\mathbb{Y}_{c,t} \in \{wL_{c,t}, rK_{c,t}, (wL_{c,t}+ rK_{c,t}), pQ_{c,t}, Q_{c,t}, p_{c,t}, LProd_{c,t}, TFP_{c,t}\}$ with $LProd_{c,t}$ and $TFP_{c,t}$ as a measure for labor productivity and total factor productivity (TFP), respectively. 
In line with equation~\eqref{equation:tax_revenues}, we control for the same set of country-level controls $Z_{c,t}$, include country and year FE, and cluster standard errors at the country and year level. Results are presented in Table \ref{tab:regr_realincome_all_9516}.

\begin{table}[H]
{\centering 
    \caption{Real income effect}
        \label{tab:regr_realincome_all_9516}

\begin{adjustbox}{width=0.9\textwidth}\begin{tabular}{l cccccccc}
\toprule \addlinespace
\csname @@input\endcsname{inputs/all_95_16/Results5RICTint_ct_All_95t16}
\bottomrule \addlinespace
\csname @@input\endcsname{inputs/all_subperiod/Results5RICTint_ct_All_95t07}
\bottomrule \addlinespace
\csname @@input\endcsname{inputs/all_subperiod/Results5RICTint_ct_All_08t16}
\bottomrule
\end{tabular}\end{adjustbox}\\ \noindent }
\scriptsize
Notes: \sym{*}\(p<0.05\), \sym{**}\(p<0.01\), \sym{***}\(p<0.001\). 
All regressions use data for nineteen European countries during 1995-2016 and include: GDP growth, government consolidated gross debt as \% of GDP; government interest payable as \% of GDP; net government lending/borrowing as \% of GDP; gross fixed capital formation as \% of GDP; period average exchange rate; and country ($c$) and year ($t$) fixed effects. Labor productivity is measured as the share of gross output volumes ($Q$) over the total number of hours worked. TFP is calculated as the residual from an OLS regression of gross output volumes ($Q$) on a translog production function including capital volumes ($K$), total number of hours worked ($L$) and intermediate input volumes ($M$). Standard errors are clustered both at the country and year level.
\end{table}

We find that robot diffusion is significantly negatively associated with incomes from capital and labor. It is also negatively associated with aggregate sales and prices. 
Prior to 2007, we observe robot diffusion to be accompanied with a contraction in output quantities, which is reversed after 2008. 
After 2008, the effects of robot diffusion on factor income and sales are insignificant. 
The depth of automation exhibits roughly the same effects on factor markets as for robots, but these are weakly significant. 
While we do not see any significant interaction between robots and our productivity measures, we observe a positive link between the automation depth and TFP, but a negative one for labor productivity. 

For ICT diffusion, we observe a positive relation of ICT with labor income and labor productivity. This is most significant pre-2007. During this period, we additionally find a weakly significant link with the value of aggregate output, while after 2008, we find a positive effect on capital income and total factor income $wL_{c,t} + rK_{c,t}$. Prior to 2007, ICT diffusion is negatively associated with TFP while this relationship is reversed post-2008.

\FloatBarrier
\section{Robustness checks}
\label{sec:robustness}

In this section we discuss the key parts of further analysis we have undertaken to ensure the robustness of our findings.\footnote{For a more detailed presentation and discussion of the results we refer the reader to the relevant section of the Online Appendix Sections \ref{SM:robustness_endogeneity}-\ref{SM:robustness_taxes}.}

\subsection{Endogeneity}
Ideally, we would like to measure the pure impact of technological progress in ATs as an exogenous driver of AT diffusion to see how it affects the economy and public revenues. However, we only observe patterns of AT adoption which are likely to endogenously depend on economic dynamics. 

To alleviate such endogeneity concerns, we employ three robustness checks that rely on lagged data and Instrumental Variables (IV). 
First, we use lagged ($t-1$) instead of contemporaneous ($t$) robot- and ICT-intensity as explanatory variables to allow for effects that may take over one period to materialize. 
Next, we use an IV approach where deeper lags, i.e. $t-1$, $t-2$ and $t-3$, instrument for the contemporaneous AT diffusion measures. The estimates 
are consistent with the baseline results. 

Furthermore, 
we apply an alternative IV approach 
inspired by \citet{blanas2019afraid} following the idea that AT imports from other countries should be driven by technological advances in ATs, but are entirely exogenous to the economic dynamics in country $c$. For this we use robot and ICT product imports by all countries in the world except $c$ as an instrument for robot and ICT diffusion in country $c$.  
Again, we obtain qualitatively consistent point estimates for the coefficients, but the validity tests indicate a weak explanatory power at the first stage. 
Summing up, the lag-based and both IV approaches to cope with the endogeneity of AT diffusion measures support our analysis qualitatively, but given the aggregation-level of the analysis the trade-based IV approach suffers from weak instruments.
\footnote{We have also experimented with alternative external IV approaches by constructing Bartik-style instruments, but we ran into similar issues in terms of the validity of our instruments. For more details see Online Appendix Section \ref{SM:robustness_endogeneity}.} 

\subsection{Further tests}
In a series of further checks we include additional controls, such as trade, corporate taxation, and distribution. For some variables, we have incomplete time and country coverage, and thus why we abstain from directly including them in our main analysis.

First, we test the sensitivity of our results against changes in the tax systems. While comprehensive data covering the whole range of different taxes is not available, we proxy tax reforms using data on corporate taxation for a smaller period but for all countries in our sample. We use two different data sources. 
First, we repeat all baseline country-level regressions and include as an additional control the corporate tax rate ($CRT_{c,t}$) sourced from KPMG. 
This data are available between 2003-2016, and thus only the results for the post-2008 period are comparable with the baseline analysis. 
Next, 
we repeat all baseline country-level regressions and include, as an additional control, the effective tax rate ($ERT_{c,t}$) sourced from Eurostat. 
The ETR is only available between 2006-2016 and again, only results for the post-2008 period are comparable with the baseline analysis.

Another concern regarding the robustness of our results may arise from the impact of trade. 
To capture the country-specific impact of trade, we repeat all baseline country-level regressions and include, as additional controls, the country-level imports ($Imports^{\% GDP}_{c,t}$) and exports ($Exports^{\% GDP}_{c,t}$) as percentage of GDP. 
Finally, to explore the nexus between distribution and taxation, we examine the progressiveness of taxation. 
To do so, we rely on the same empirical specification used to understand the determinants of taxation, but now our regressions include, as an additional control, the Gini coefficient measuring cross-industry wage inequality ($Gini^{w}_{c,t}$) sourced from Eurostat. 

Overall, the results from these exercises are qualitatively consistent with our main findings, albeit in some cases of lower statistical significance which might be due to differences in the data coverage.\footnote{For a detailed presentation of the results and data used see Online Appendix Section \ref{SM:robustness_taxes}.}

\FloatBarrier
\section{Discussion}
\label{sec:discussion}
\subsection{Answering the research questions} 
Now, we return to the research questions outlined in Section \ref{sec:introduction}: 
\begin{enumerate}
	\item \emph{What is the effect of automation on aggregate tax revenues at the country level in absolute terms and relative to GDP?}
	\item \emph{What is the relationship between AT diffusion and the composition of taxes by source, i.e. labor, capital and goods?}
	\item \emph{How can these effects be traced back to the three effects through which automation affects economic production?}
\end{enumerate}	
We have shown that it is important to be specific when talking about the impact of automation on the economy and fiscal revenues as robots and ICT are not only conceptually but also economically different technology types. 
	
Industrial robots as pure automation technology exhibit a negative relationship with aggregate tax revenues. This effect is strongest in the period until 2007 when robot diffusion was accompanied by a negative real income effect, i.e. declining factor revenues and output at the aggregate level. 
Our results suggest that robots are labor-replacing in automation-intensive industries, but not for any impact on labor demand at the country-level, though we find a negative association between robot diffusion and country-level wages. 
At this time, aggregate tax revenues decreased which was driven by lower tax revenues raised from capital. A shift from capital taxation towards the taxation of goods occurred.

After 2008, the impact of robots disappears: while industrial robots are still labor-replacing in automating industries, we cannot find any significant effect on aggregate factor earnings, but we find prices for final goods to decrease and aggregate demand in volumes to rise. Aggregate consumption in values is not significantly affected and the same holds for taxation: none of our taxation measures is significantly impacted during this period. 
Looking at the full time period, robot diffusion was accompanied by a shift from labor taxation towards taxes on goods. 

The relationship between ICT diffusion and taxation is different. 
For the full sample, we find a negative impact of ICT on aggregate tax revenues, taxes from capital, and a shift from the taxation of capital to the taxation of goods. These effects are not significant for any of the sub-periods, but qualitatively consistent during the pre-2007 phase. 

Again, the economic effects differ across periods. 
Prior to 2007, ICT diffusion shows a positive association with labor demand and labor income in automating industries. At the country level, ICTs exhibit a positive overall effect on labor income, reflected in wage increases along with declines in employment. 
Our analysis shows that rising labor income is a key determinant of tax revenues. Hence, we expect a technology that raises labor income also induces increases in tax revenues. Against our intuition, the impact of ICT on taxation is negative and disappears during the second period. 
We observe no effect on country-level capital income. We also find evidence for a weak increase in aggregate demand. These effects taken together offer an explanation for the shift from capital taxes towards an increasing relative importance of taxes on goods. 
Post-2008, we do not observe any significant relationship between ICT and labor income, but a weak positive effect on capital income. During that period, the impact on taxation becomes insignificant. 

The effects of ICT on productivity are ambiguous with an increasing labor productivity, but decreasing TFP in the pre-2007 period. After 2008, the impact on TFP is significantly positive. 
Technology diffusion is an inherently dynamic process with adoption lags, learning, creative destruction, and sometimes it takes time until the economic benefits of technological advance unfold. 
The differential findings across periods highlight that it may be insufficient to focus on a short time period when studying the impact of ATs on the economy and fiscal revenues. 


Note that the sample size for the post-2008 period is smaller, and thus could explain the lower significance of the post-2008 findings. In addition, the period after the financial crisis might exhibit distorting patterns of production and fiscal policies that potentially undermine the capacity to capture the impact of technological change. 

\subsection{Challenges of taxation analysis}
We need to emphasize a few challenges that we face in our work. First, tax systems are complex and have been subject to reform policies prior and post to the financial crisis. The financial crisis in 2008 was a key driver of structural reforms. 
These policy changes are difficult to control for, especially in a set of heterogeneous countries with diverse cultures of taxation that evolved differently over decades. We cope with these peculiarities by splitting our analysis into different periods of time and conducting a battery of checks to exclude a range of potential confounding factors.

A second challenge related to the tax data is the notion of endogeneity, where two types of endogeneity might be relevant. First, we do not know to what extent automation and its economic impact are affected by particular tax rules.
We cope with this problem through a series of robustness checks using data on corporate tax rates as additional controls and find that this does not affect our results. Moreover, checking country-level time series data on implicit taxes on labor and capital, we do not observe any distortionary effects, which differs from observations for the US \citep{acemoglu2020does}. 
In our sample, there is no ex-ante reason that variations in AT diffusion in Europe can be attributed to changes in relative taxation. 

A second concern about endogeneity arises due to the cyclicality of investment decisions. In particular, we do not know whether we observe the impact of AT diffusion on the economy or vice versa. We cope with this through a series of robustness checks using lagged instead of contemporaneous AT diffusion in the regression and through two types of IV approaches relying on lagged AT and trade data. We find that our observations are robust against these alternative specifications. 

Finally, it should also be noted that our analysis only briefly touched upon distributional effects. Conceptually, we have implicitly assumed a linear relationship between country-level wage and capital income, consumption and taxation. However, households with different income levels consume and save differently, and employees earning different wages face different tax rates dependent on the progressiveness of taxation. Our results suggest that automation may increase cross-industry wage inequality. Inequality is a major issue in the literature on automation, but an in-depth analysis in this context is left for future study.

\section{Conclusion}
\label{sec:conlusion}
In this study, we explore the effects of automation on taxation. We introduce a stylized theoretical framework that decomposes tax revenues into three broader sources of taxation distinguishing between taxes on labor, capital, and consumption. We link these sources to production and identify three economic effects of automation. This provides the theoretical basis to draw a link from micro-level AT adoption to aggregate flows of tax revenues.
In contrast to existing studies on automation and taxation, we do not look at the impact of taxes on firms' adoption decisions. Instead, we are the first to study empirically the relationship between automation and tax revenues taking adoption decisions as given. 

Our study suggests that the nexus between taxation and automation is complex and requires a careful monitoring of the economic side effects of technological change. Preceding studies argued that policy makers should be concerned about the sustainability of public finances when intelligent machinery replaces labor and undermines the basis of taxation. Our study is the first to explore the empirical basis of this claim. Overall, our findings suggest that there is no empirical evidence supporting that tax revenues are negatively affected by ATs in the long run. Specifically, whether automation erodes taxation depends on the technology and the stage of diffusion, and concerns about public budgets might be short sighted when focusing on the short term and ignoring other technology trends.

\newpage
\begin{spacing}{1}
\putbib 
\end{spacing}
\end{bibunit}

\begin{bibunit}

\maketitle
\thispagestyle{empty}

\newpage
\pagenumbering{arabic}
\setcounter{page}{1} 
\onehalfspacing

\newpage
\FloatBarrier
\appendix
\renewcommand{\appendixname}{Appendix}
\renewcommand{\thesection}{\Alph{section}} \setcounter{section}{0}
\renewcommand{\thefigure}{\Alph{section}.\arabic{figure}} \setcounter{figure}{0}
\renewcommand{\thetable}{\Alph{section}.\arabic{table}} \setcounter{table}{0}
\renewcommand{\theequation}{\Alph{section}.\arabic{table}} \setcounter{equation}{0}


\FloatBarrier
\section{Data construction}
\label{appendix:data}


Overall, we combine different data sets at different aggregation levels with varying coverage in terms of countries, industries, and time. After merging the data as described below, we end up with two samples. 
The first sample is a country-level panel data set covering the whole economy, from agriculture to public sectors, for nineteen European countries during 1995-2016.\footnote{The sample is unbalanced since data are missing for: Lithuania, Latvia, and the United Kingdom during the base year, i.e. 1995; and Denmark, Portugal, Slovenia and Slovakia for the period 1995-1999.}

The second sample is an industry-level panel data set covering only automation-intensive industries. We classify industries as automation-intensive when information on the use of robots exists, since the data coverage of industries is endogenous, i.e. only significant customers of industrial robots are reported \citep{IFR2020report}.
These data cover the same set of countries and years as the country-level data excluding Portugal due to missing information. The industries classified as automation-intensives include: agriculture; mining and quarrying; ten manufacturing industries; electricity, gas, and water supply; construction; and education, research and development.\footnote{The sample is unbalanced since certain country-industry-year combinations are missing. Generally, the industry and year coverage is rather limited for Eastern European countries i.e. Estonia, Lithuania, Latvia, Slovenia and Slovakia. Details on the coverage are provided in the Appendix Table \ref{tab:coverage_cit}.}

\subsection{Sources of tax revenue}
\label{sec:empirical_approach_tax}
Taxes are part of our country-level data and compiled on the basis of the Global Revenue Statistics Database of the \citet{OECD2020data}. We use the OECD terminology to define the tax aggregates as follows:
\begin{itemize}
	\item $T_{c,t}^l$---taxes on labor are the sum of \emph{Social security contributions (2000)} and \emph{Taxes on payroll and workforce (3000)};   
	\item $T_{c,t}^k$---taxes on capital are the sum of \emph{Taxes on income, profits and capital gains (1000)} and \emph{Taxes on property (4000)};\footnote{We include property taxes as part of capital taxes because: (1) they consist largely of taxes on corporate property; and (2) we interpret property as part of the productive capital that is used to provide economic services to final consumers. This interpretation also holds for the majority of private property taxes. For example, taxes on houses are one of the most significant parts of property taxes. Housing is a service consumed by households even if private housing is not traded on the market. This interpretation is not applicable to other components of property taxes (e.g. taxes on gifts). However, tax revenues raised from these residual accounts are negligibly small. The total block of property taxes accounts on average for less than $2\%$ of GDP. Checks excluding all \textit{4000}-tax codes confirms that this does not alter the results.}
	\item and $T_{c,t}^y$---taxes on goods given by \emph{Taxes on goods and services (\textit{5000})},
\end{itemize}
where the numbers in parentheses indicate the tax code from the OECD tax classification system \cite[][A.1]{OECD2019}.\footnote{
In this analysis, we ignore residual taxes (\textit{6000}) which, on average across OECD countries, account for approximately 0.2\% of GDP and 0.6\% of total taxation.} 

To describe the impact on tax revenues, we use taxes measured in national currency. To put this in relation to production, we look at taxes measured as percentage of GDP. 
For an analysis on the structure of taxation, we use tax data measured as percentage of total taxation.

\subsection{Economic variables}
\label{sec:empirical_approach_economic_vars}
Empirical proxies for factor income and consumption at the country level are aggregates of NACE Rev. 2 (ISIC Rev. 4) industry-level data from the EUKLEMS database \citep{EUKLEMS2019data, stehrer2019industry, adarov2019tangible}. 
We use:
\begin{itemize}
	\item $LAB$ for $w_{c,t} L_{c,t} = \sum_{i \in I_c} w_{i,c,t}L_{i,c,t}$;   
	\item $CAP$ for $r_{c,t} K_{c,t} = \sum_{i \in I_c} r_{i,c,t}K_{i,c,t}$;
	\item and $GO$ for $p_{c,t}Q_{c,t} = \sum_{i \in I_c} p_{i,c,t}Q_{i,c,t}$,\footnote{$LAB$ is computed as the compensation of employees in current prices of national currency in  million times the ratio of total hours worked by persons engaged over total hours worked by employees, which assumes that in each industry the self-employed receive the same hourly wage as the employees. $CAP$ is the capital compensation calculated as the value added minus labor compensation. Note that we use the value of the capital stock as proxy for the rate of return to capital. $GO$ is the gross output in current prices of national currency in  million.}
\end{itemize}
where $w_{i,c,t}$, $r_{i,c,t}$ and $p_{i,c,t}$, are computed by dividing the respective variables measured in values to their volumes.\footnote{Specifically, we source from EUKLEMS $L_{c,t} = \sum_{i \in I_c} L_{i,c,t}$, $K_{c,t} = \sum_{i \in I_c} K_{i,c,t}$ and $Q_{c,t} = \sum_{i \in I_c} Q_{i,c,t}$ as the number of hours worked in million ($H\_EMPE$), the net capital stock volume of all assets in million ($Kq\_GFCF$), and the gross output volume in million ($GO\_Q$). Similarly, we construct country-level $w_{c,t}$, $r_{c,t}$ and $p_{c,t}$ by dividing the corresponding country-level aggregates in values by volumes $L_{c,t}$, $K_{c,t}$ and $Q_{c,t}$, respectively.} 

Automation may lead to industrial restructuring. To measure this, we construct two structural indicators using industry-level data. First, we compute the service sector market share: $Services_{c,t} = \tfrac{\sum_{i \in I^s_{c}} p_{i,c,t}Q_{i,c,t}}{\sum_{i \in I_c} p_{i,c,t}Q_{i,c,t}}$, where $I^s_{c}$ is the set of service industries in $c$.\footnote{We define service industries as NACE Rev. 2 (ISIC Rev. 4) 2-digit codes 45-99 or 1-digit codes G-U.} 
Second, we compute as a measure of industrial concentration the Hirschmann-Herfindahl index on the basis of industry shares in total production, i.e. $HHI^{}_{c,t} = \sum_{i \in I_c} \left( \tfrac{p_{i,c,t}Q_{i,c,t}}{p_{c,t}Q_{c,t}} \right)^2$. 

For an indicator of cross-industrial wage inequality, we use industry-level data on wages to calculate the country-level Gini coefficient as follows: $Gini^{w}_{c,t} = \frac{\sum_{i=1}^{I_c}(2i-I_c-1)w_{i,c,t}}{I_c \sum_{i=1}^{I_c}w_{i,c,t}}$, where $I_c$ are all industries in country $c$ and $i$ is now the rank of industry-level wages in ascending order. Analogously, we compute $Gini^L_{c,t}$ which measures the distribution of employment across industries. A higher level of $Gini^{w}_{c,t}$ ($Gini^{L}_{c,t}$) indicates a more unequal distribution of wage (labor) across industries.

To examine the impact of automation on productivity, we use industry-level data to calculate labor productivity $LProd_{c,t}$ as the share of gross output volumes over the total number of hours worked. We also estimate total factor productivity $TFP_{c,t}$ as the residual from an OLS regression of gross output volumes on a translog production function of volumes of capital, labor (hours worked) and material inputs \cite[cf.][]{stehrer2019industry}. 

In the tax regressions, we additionally control for determinants of taxation identified in the literature \cite[e.g.][]{castro2014determinants, castaneda2018tax}.
We include GDP growth and different indicators of public finances sourced from \citet{Eurostat2020data}, such as: government consolidated gross debt as \% of GDP ($Debt^{\%GDP}_{c,t}$); net government lending/borrowing as \% of GDP ($Lending^{\%GDP}_{c,t}$); government interest payments on debt as \% of GDP ($Interest_{c,t}^{\%GDP}$); and public gross fixed capital formation as \% of GDP ($GovInv^{\%GDP}_{c,t}$). We capture the role of trade by including the period average exchange rate ($XRate_{c,t}$) from the \citet{OECD2020data} data set. 
Robustness checks including additional controls, such as corporate tax rates, and import and export rates are provided in \ref{SM:robustness_trade}.

\subsection{Measuring automation}
\label{sec:empirical_approach_ai_data}
We use two measures for automation calculated on the basis of: (1) the number of operational industrial robots; and (2) the capital stock of ICT, including computer software and databases. 

The data on industrial robots is from the International Federation of Robotics (IFR) \citep{IFR2020data}. An industrial robot is defined as {``automatically controlled, reprogrammable, multipurpose manipulator [...] for industrial applications''} \citep{IFR2020report}.\footnote{This definition follows the ISO norm 8373 (see \url{https://www.iso.org/obp/ui/\#iso:std:iso:8373:ed-2:v1:en}).}
The IFR provides data on deliveries and stocks of industrial robots at the industry level. Industrial robots are a measure of automation because they can readily replace humans in the execution of specific tasks \cite[see][]{graetz2018robots, de2020rise, acemoglu2020robots, faber2020robots}. 

To measure the extent to which robots became part of an industry's production technology, we construct the \emph{robot density} measure as the number of operational robots over the number of hours worked by human labor in industry $i$, i.e. $R_{i,c,t} = \tfrac{\#Robots_{i,c,t}}{L_{i,c,t}}$. For the country level analysis, we compute $R_{c,t} = \tfrac{\sum_{i \in I_{c}} \#Robots_{i,c,t}}{\sum_{i \in I_c} L_{i,c,t}}$.\footnote{For the construction of the number of operational robots in each industry ($\#Robots_{i,c,t}$), we closely follow \cite{graetz2018robots}.}

As a second automation indicator, we use the \emph{ICT-intensity} measured as net ICT capital stock per hour worked $L_{i,c,t}$.
The data on ICT capital is taken from \citet{EUKLEMS2019data} as the sum of net capital stock volumes of computing equipment $(Kq\_IT$), communications equipment $(Kq\_CT)$, and computer software and databases $(Kq\_Soft\_DB)$. It includes both tangible (hardware) and intangible (data bases and software) ICTs.

The coverage of industries differs for data on robots and ICT. Data on ICT covers the whole economy, except for all industries in Portugal and certain industries and/or years in Eastern European countries. Robot data are also available for more countries than those in the ICT data set, but reported only for the following industries: agriculture; mining and quarrying; ten manufacturing industry groups; electricity, gas and water supply; construction; and education, research and development (see Appendix Table \ref{tab:coverage_cit}).

We include these two types of automation to potentially account for two different AT types. 
Robots and ICTs can be distinguished by the type of task they can execute: robots are designed to perform manual tasks, while ICT has a stronger link to cognitive tasks. 
While robots are pure ATs that execute a clearly defined task previously performed by humans, it is less clear whether this also applies to ICTs. ICTs can be flexibly applied for many different tasks and, to some extent, these tasks do not have a clear analogue in the range of tasks that can be executed by humans. 

In our analysis, we introduce both diffusion measures simultaneously and as an interaction term. Robot-ICT interaction captures complementarities between the two ATs or otherwise stated the depth of automation, i.e. the extent to which both manual and cognitive tasks are performed by machinery. Concerns about multicollinearity can be ruled out since we find a very low correlation between both measures (with a correlation coefficient of $22\%$ for all countries in the sample and with $34\%$, $19\%$, and $63\%$ for Eastern, Northern and Southern European countries, respectively. 
The bottom left panel in Figure \ref{fig:key_vars_all_regions} shows a time series plot of the z-score normalized data on robot and ICT diffusion. The figure shows that the pace of robot diffusion increased sharply in the period after 2008.

A major concern is over the potential endogeneity of the diffusion measures. While we study the impact of AT diffusion on the economy, the causality can run vice-versa, and thus AT adoption could be contingent on economic dynamics. For instance, a well documented macroeconomic regularity is that investment cycles are positively correlated with cyclical boosts and busts in the economy \citep{stock1999business, anzoategui2019endogenous}. 
In our analyses, while we control for gross capital formation which captures the cyclicality of general investments, AT-specific investments could still follow a different trend.
To cope with endogeneity of this sort, we apply a series of robustness checks, including the use of lagged ($t-1$) diffusion measures and exploring two types of IV regressions. For the latter, we experiment with: (1) an internal IV strategy using deeper lags (up to $t-3$) of the AT measures as instruments; and (2) an external IV strategy, whereby we use data on global trade in robots and ICT products to construct external instruments for AT adoption (see \ref{SM:robustness_endogeneity}).

\newpage
\section{Additional Tables and Figures}

\begin{table}[H] 
{\centering
\caption{List of NACE Rev.2 (ISIC Rev.4) industry groups in industry level data.}
\label{tab:industry_codes}
\begin{tabular}{lcc} \toprule
\multicolumn{2}{c}{Industry aggregation:} & \\ \cmidrule(l){1-2}
EUKLEMS & IFR  & Description of industries in IFR dataset \\ \midrule
01t03 & 01t03 & A-B-Agriculture, forestry, fishing \\
05t09 & 05t09 & C-Mining and quarrying \\
10t12 & 10t12 & 10-12-Food and beverages \\
13t15 & 13t15 & 13-15-Textiles \\
16t18 & 16 & 16-Wood and furniture \\
16t18 & 17t18 & 17-18-Paper \\
19t21 & 19t20 & 20-21-other chemical products n.e.c. \\
19t21 & 21 & 19-Pharmaceuticals, cosmetics \\
22t23 & 22 & 22-Rubber and plastic products (non-automotive) \\
22t23 & 23 & 23-Glass, ceramics, stone, mineral products (non-auto \\
24t25 & 24 & 24-Basic metals \\
24t25 & 25 & 25-Metal products (non-automotive) \\
26t27 & 26t27 & 26-27-Electrical/electronics \\
28 & 28 & 28-Industrial machinery \\
29t30 & 29 & 29-Automotive \\
29t30 & 30 & 30-Other vehicles \\
31t33 & 32 & 91-All other manufacturing branches \\
35t39 & 35t39 & E-Electricity, gas, water supply \\
41t43 & 41t43 & F-Construction \\
85 & 85 & P-Education/research/development \\
 Rest & Rest & 90-All other non-manufacturing branches \\ \bottomrule
\end{tabular}
\\ }
\scriptsize
Notes: EUKLEMS and IFR refer to the aggregation of NACE Rev.2 (ISIC Rev.4) 2-digit industries considered in the EUKLEMS and IFR data set, respectively. The industry level analysis in this paper is based on the more aggregate EUKLEMS industry aggregation.
\end{table}

\begin{landscape}
\begin{table}[H]
\caption{Time period coverage of each country industry pair in the industry level sample.}
\label{tab:coverage_cit}
{\centering
\begin{adjustbox}{width=1.4\textwidth}
\begin{tabular}{lccccccccccccccc} \toprule

&\multicolumn{15}{c}{Industry groups based on ISIC Rev.4 (NACE Rev.2) 2-digit codes used in the industry-level sample} \\ \cmidrule(l){2-16}
Country & 01t03 & 05t09 & 10t12 & 13t15 & 16t18 & 19t21 & 22t23 & 24t25 & 26t27 & 28 & 29t30 & 31t33 & 35t39 & 41t43 & 85 \\ \midrule
AT & 1995-2016 & 1995-2016 & 1995-2016 & 1995-2016 & 1995-2016 & 1995-2016 & 1995-2016 & 1995-2016 & 1995-2016 & 1995-2016 & 1995-2016 & 1995-2016 & 1995-2016 & 1995-2016 & 1995-2016 \\
BE & 1995-2016 & 1995-2016 & 1995-2016 & 1995-2016 & 1995-2016 & 1995-2016 & 1995-2016 & 1995-2016 & 1995-2016 & 1995-2016 & 1995-2016 & 1995-2016 & 1995-2016 & 1995-2016 & 1995-2016 \\
CZ & 1995-2016 & 1995-2016 & 1995-2016 & 1995-2016 & 1995-2016 & 1995-2016 & 1995-2016 & 1995-2016 & 1995-2016 & 1995-2016 & 1995-2016 & 1995-2016 & 1995-2016 & 1995-2016 & 1995-2016 \\
DE & 1995-2016 & 1995-2016 & 1995-2016 & 1995-2016 & 1995-2016 & 1995-2016 & 1995-2016 & 1995-2016 & 1995-2016 & 1995-2016 & 1995-2016 & 1995-2016 & 1995-2016 & 1995-2016 & 1995-2016 \\
DK & 1995-2016 & 1995-2016 & 1995-2016 & 1995-2016 & 1995-2016 & 1995-2016 & 1995-2016 & 1995-2016 & 1995-2016 & 1995-2016 & 1995-2016 & 1995-2016 & 1995-2016 & 1995-2016 & 1995-2016 \\
EE & 2000-2016 & 2000-2016 & N/A & N/A & N/A & N/A & N/A & N/A & N/A & N/A & N/A & N/A & 2000-2016 & 2000-2016 & 2000-2016 \\
ES & 1995-2016 & 1995-2016 & 1995-2016 & 1995-2016 & 1995-2016 & 1995-2016 & 1995-2016 & 1995-2016 & 1995-2016 & 1995-2016 & 1995-2016 & 1995-2016 & 1995-2016 & 1995-2016 & 1995-2016 \\
FI & 1995-2016 & 1995-2016 & 1995-2016 & 1995-2016 & 1995-2016 & 1995-2016 & 1995-2016 & 1995-2016 & 1995-2016 & 1995-2016 & 1995-2016 & 1995-2016 & 1995-2016 & 1995-2016 & 1995-2016 \\
FR & 1995-2016 & 1995-2016 & 1995-2016 & 1995-2016 & 1995-2016 & 1995-2016 & 1995-2016 & 1995-2016 & 1995-2016 & 1995-2016 & 1995-2016 & 1995-2016 & 1995-2016 & 1995-2016 & 1995-2016 \\
GR & 1995-2016 & 1995-2016 & 1995-2016 & 1995-2016 & 1995-2016 & 1995-2016 & 1995-2016 & 1995-2016 & 1995-2016 & 1995-2016 & 1995-2016 & 1995-2016 & 1995-2016 & 1995-2016 & 1995-2016 \\
IT & 1995-2016 & 1995-2016 & 1995-2016 & 1995-2016 & 1995-2016 & 1995-2016 & 1995-2016 & 1995-2016 & 1995-2016 & 1995-2016 & 1995-2016 & 1995-2016 & 1995-2016 & 1995-2016 & 1995-2016 \\
LT & 1995-2016 & 1995-2016 & N/A & N/A & N/A & N/A & N/A & N/A & N/A & N/A & N/A & N/A & 1995-2016 & 1995-2016 & 1995-2016 \\
LV & 1995-2016 & 2000-2016 & N/A & N/A & N/A & N/A & N/A & N/A & N/A & N/A & N/A & N/A & 2000-2016 & 1995-2016 & 2000-2016 \\
NL & 1995-2016 & 1995-2016 & 1995-2016 & 1995-2016 & 1995-2016 & 1995-2016 & 1995-2016 & 1995-2016 & 1995-2016 & 1995-2016 & 1995-2016 & 1995-2016 & 1995-2016 & 1995-2016 & 1995-2016 \\
SE & 1995-2016 & 1995-2016 & 1995-2016 & 1995-2016 & 1995-2016 & 1995-2016 & 1995-2016 & 1995-2016 & 1995-2016 & 1995-2016 & 1995-2016 & 1995-2016 & 1995-2016 & 1995-2016 & 1995-2016 \\
SI & 2000-2016 & 2000-2016 & N/A & N/A & N/A & N/A & N/A & N/A & N/A & N/A & N/A & N/A & 2000-2016 & 2000-2016 & 2000-2016 \\
SK & 2000-2016 & 2000-2016 & 2000-2016 & 2000-2016 & 2000-2016 & 2000-2016 & 2000-2016 & 2000-2016 & 2000-2016 & 2000-2016 & 2000-2016 & 2000-2016 & 2000-2016 & 2000-2016 & 2000-2016 \\
 UK & 2007-2016 & 1995-2016 & 1995-2016 & 1995-2016 & 1995-2016 & 1995-2016 & 1995-2016 & 1995-2016 & 1995-2016 & 1995-2016 & 1995-2016 & 1995-2016 & 1995-2016 & 1995-2016 & 1995-2016 \\ \bottomrule
\end{tabular}

\end{adjustbox} \\}
\scriptsize
Notes: This table presents the year coverage across countries and industries for the industry level sample used in the analysis. Industries refer to groupings of NACE Rev.2 (ISIC Rev.4) 2-digit industry codes and are discussed in detail in Table~\ref{tab:industry_codes}. 
\end{table}

\begin{table}[H]
\caption{Descriptive statistics}
\label{tab:descriptives_all}
\scriptsize
{\centering
\begin{adjustbox}{width=1.4\textheight}
\begin{tabular}{l cccc| ccc| ccc | c |c|c|cccc|cc}
\toprule
& \multicolumn{4}{c}{\% of GDP} &  \multicolumn{3}{c}{\% of total tax} & \multicolumn{3}{c}{Production} & \multicolumn{1}{c}{GDP} & \multicolumn{1}{c}{Services} & \multicolumn{1}{c}{} & \multicolumn{4}{c}{\% of GDP}& \multicolumn{2}{c}{Gini}\\ \cmidrule(l){2-5} \cmidrule(l){6-8} \cmidrule(l){9-11}  \cmidrule(l){15-18} \cmidrule(l){19-20}
& $T$ & $T^l$ & $T^k$ & $T^y$ & $T^l$ & $T^k$ & $T^y$ & $wL$ & $rK$ & $pQ$ & $growth$  & ${pQ}$ & $HHI$ & $Debt$ & $Interest$ & $Lending$ & $GovInv$&$w$&$L$  \\\midrule
\csname @@input\endcsname{inputs/descriptives/Sumstat_controls_ct_All}
\end{tabular}
\end{adjustbox} \\}
\scriptsize
Notes: This table shows the main descriptives (mean, standard deviation, minimum, median, maximum) of the core variables included in the regression analyses covering all nineteen European countries during the period 1995-2016. Further information about the data is provided in the main article (see Section \ref{sec:empirical_approach}). 
\end{table}

\end{landscape}

\newpage
\FloatBarrier

\begin{figure}[H]
{\centering
\caption{Evolution of implicit taxes on labor and capital}
\label{fig:rel_taxes_c}	    
\includegraphics[scale=1.1]{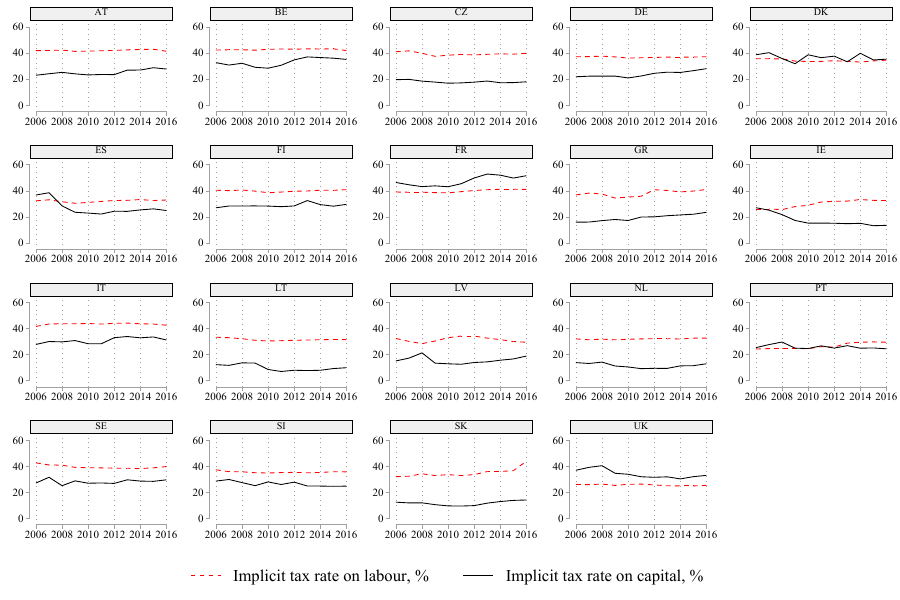} \\}
\scriptsize
Notes: These figures show the evolution of implicit tax rates on labor and capital in whole Europe and for the subsets of Eastern Northern and Southern European countries as defined in the text. The data are downloaded from the European Commissions tax database \citep{EC2021taxdata}. See also \citet{EC2020taxreport}. 
\end{figure}

\newpage
\section{Further determinants of taxation}
\label{appendix:add_results_determinants_taxation}
Here, we briefly outline other determinants of taxation that are included in the regressions but not discussed in the main text. These regressors are motivated by the literature on taxation and cover: GDP growth; the market share of services $Services_{c,t}$; public finance related indicators (debt, interest payments and deficit); measures for industrial concentration $HHI_{c,t}$; the exchange rate $XRate_{c,t}$ (i.e. US\$ per Euro) as proxy for trade; and public investments $GovInvest^{\%GDP}_{c,t}$. 

Labor taxes in absolute terms, in relation to GDP and as share in total taxation exhibit a positive correlation with the service share. 
This pattern holds for both sub-periods pre- and post-2008. 
High indebtedness and higher deficits are positively related to taxes in absolute terms and measured in percentage GDP. 
Net lending as percentage GDP is positively correlated with all taxes except from labor taxes which are negatively related to deficits. 

We find a higher exchange rate $XRate_{c,t}$ to be negatively related to taxes on capital, goods and in total, measured in absolute terms and as percentage of GDP. We also observe a higher exchange rate to be positively related to the relative tax contribution of labor at the cost of taxes on capital.
\footnote{Note that the exchange rate varies across European countries only in the time dimension since the EU's Exchange Rate Mechanism aims to keep exchange rate fluctuations between the Euro and other European currencies flat (see also \url{https://stats.oecd.org/glossary/detail.asp?ID=3055}). Hence, $XRate_{c,t}$ captures the competitiveness of European countries on global markets but can not be interpreted as an indicator for within-European trade.}

\begin{table}[H]
{\centering
\caption{Taxation and the structure of production}
\label{tab:regr_pre_tax_prod6}
\begin{adjustbox}{width=\textwidth}\begin{tabular}{l cccc|cccc|ccc}
\toprule \addlinespace
\csname @@input\endcsname{inputs/all_95_16/Results2RICTint_ct_All_95t16}
\bottomrule
\end{tabular}\end{adjustbox} \\ } \scriptsize
Notes: \sym{*}\(p<0.05\), \sym{**}\(p<0.01\), \sym{***}\(p<0.001\). Regression results to examine the link between tax aggregation and economic production for nineteen European countries during the period 1995-2016. All regressions use country level data and include: GDP growth ($GDPgrowth_{c,t}$); share of gross output produced in service industries ($Services_{c,t}$); Herfindahl-Hirschman Index based on the gross output shares of macro-sectors ($HHI_{c,t}$); government consolidated gross debt as \% of GDP ($Debt^{\%GDP}_{c,t}$); government interest payable as \% of GDP ($Interest^{\%GDP}_{c,t}$); net government lending/borrowing as \% of GDP ($Lending^{\%GDP}_{c,t}$); gross fixed capital formation as \% of GDP ($GovInv^{\%GDP}_{c,t}$); period average exchange rate ($XRate_{c,t}$); and country ($c$) and year ($t$) fixed effects. For the first block (Taxes in ln of national currency), $wL_{c,t}$, $rK_{c,t}$ and $pQ_{c,t}$ are expressed as the natural logarithm ($ln$) while for the last two blocks they are expressed as \% of GDP. Standard errors are clustered both at the country and year level.
\end{table}
\newpage

\begin{table}[!hb]
{\centering 
\caption{Taxation and the structure of production before 2007}
\begin{adjustbox}{width=\textwidth}\begin{tabular}{l cccc|cccc|ccc}
\toprule \addlinespace
\csname @@input\endcsname{inputs/all_subperiod/Results2RICTint_ct_All_95t07}
\bottomrule
\end{tabular}\end{adjustbox} \\ }
    \scriptsize
Notes: \sym{*}\(p<0.05\), \sym{**}\(p<0.01\), \sym{***}\(p<0.001\). Regression results to examine the link between tax aggregation and economic production  for nineteen European countries during the period 1995-2007. All regressions use country level data and include: GDP growth ($GDPgrowth_{c,t}$); share of gross output produced in service industries ($Services_{c,t}$); Herfindahl-Hirschman Index based on the gross output shares of macro-sectors ($HHI_{c,t}$); government consolidated gross debt as \% of GDP ($Debt^{\%GDP}_{c,t}$); government interest payable as \% of GDP ($Interest^{\%GDP}_{c,t}$); net government lending/borrowing as \% of GDP ($Lending^{\%GDP}_{c,t}$); gross fixed capital formation as \% of GDP ($GovInv^{\%GDP}_{c,t}$); period average exchange rate ($XRate_{c,t}$); and country ($c$) and year ($t$) fixed effects. For the first block, $wL_{c,t}$, $rK_{c,t}$ and $pQ_{c,t}$ are expressed as the natural logarithm ($ln$) while for the last two blocks they are expressed as \% of GDP.
\end{table}

\newpage
\begin{table}[H]
{\centering 
\caption{Taxation and the structure of production after 2008}
\begin{adjustbox}{width=\textwidth}\begin{tabular}{l cccc|cccc|ccc}
\toprule \addlinespace
\csname @@input\endcsname{inputs/all_subperiod/Results2RICTint_ct_All_08t16}
\bottomrule
\end{tabular}\end{adjustbox} \\ }
    \scriptsize
Notes: \sym{*}\(p<0.05\), \sym{**}\(p<0.01\), \sym{***}\(p<0.001\). Regression results to examine the link between tax aggregation and economic production for nineteen European countries during the period 2008-2016. All regressions use country level data and include: GDP growth ($GDPgrowth_{c,t}$); share of gross output produced in service industries ($Services_{c,t}$); Herfindahl-Hirschman Index based on the gross output shares of macro-sectors ($HHI_{c,t}$); government consolidated gross debt as \% of GDP ($Debt^{\%GDP}_{c,t}$); government interest payable as \% of GDP ($Interest^{\%GDP}_{c,t}$); net government lending/borrowing as \% of GDP ($Lending^{\%GDP}_{c,t}$); gross fixed capital formation as \% of GDP ($GovInv^{\%GDP}_{c,t}$); period average exchange rate ($XRate_{c,t}$); and country ($c$) and year ($t$) fixed effects. For the first block, $wL_{c,t}$, $rK_{c,t}$ and $pQ_{c,t}$ are expressed as the natural logarithm ($ln$) while for the last two blocks they are expressed as \% of GDP.
\end{table}

\newpage
\section{Coping with endogenous automation}
\label{SM:robustness_endogeneity}
As explained in Section \ref{sec:empirical_approach_ai_data}, one major concern to the robustness of our results arises from potential endogeneity of AT diffusion.  Ideally, we would like to measure the pure impact of technological progress in ATs as exogenous driver of AT diffusion to see how technological change affects the economy and public revenues. But we can only observe patterns of AT adoption which can endogenously dependent on economic dynamics. 

To cope with these concerns, we use three different types of robustness checks relying on lagged data and IVs. Here, we present the results of these robustness checks. 

In Section \ref{SM:robustness_lagged_controls}, we used robot- and ICT-intensity from $t-1$ instead of contemporaneous diffusion measures as explanatory variable. 
In Section \ref{SM:lagged_iv}, we used an IV approach. In particular, deeper lags from $t-1$, $t-2$ and $t-3$ are used as explanatory variables on the first stage to instrument contemporaneous AT diffusion. Below the tables with the results, we report statistics to test for under-, weak- and over-identification of the IV approach. The estimates of the coefficients of this approach are consistent with the results presented in the main text. 

Another IV approach is shown in Section \ref{SM:robustness_IV}. Here, we used robot and ICT products imports by all countries in the world except $c$ as in instrument for robot and ICT diffusion in country $c$.  This approach is inspired by \citet{blanas2019afraid} and follows the idea that AT imports to other countries should be driven by technological advances in ATs, but are entirely exogenous from the economic dynamics in country $c$. 
Again, we obtain qualitatively consistent point estimates for the coefficients, but the validity tests indicate a weak explanatory power at the first stage. 

Summing up, the lag-based and both IV approaches to cope with the endogeneity of AT diffusion measures support our analysis qualitatively, but the trade-based IV approach suffers from the weakness of instruments.\footnote{We have also experimented with alternative external IV approaches to construct Bartik-style instruments, but we ran into similar issues in terms of the validity of our instruments.}

\newpage
\subsection{Using lagged ($t-1$) measures of ATs as controls}
\label{SM:robustness_lagged_controls}

\begin{table}[!h]
\caption{Taxation and automation}
{\centering
\begin{adjustbox}{width=\textwidth}\begin{tabular}{l cccc|cccc|ccc}
\toprule \addlinespace
\csname @@input\endcsname{inputs/robustness/lagged_controls/Results2shortV1_RICTint_ct_All_95t16}
\bottomrule
\end{tabular}\end{adjustbox}} \\ 
\label{tab:regr_pre_tax_robots_lagged_controls} \scriptsize
Notes: $\ast$ \(p<0.05\), $\ast\ast$ \(p<0.01\), $\ast\ast\ast$ \(p<0.001\). Regression results of aggregate flows of tax revenues on different automation measures for nienteen European countries during the period 2006-2016. All regressions use country level data and include: GDP growth, gross output share of the service sector in total economy; Herfindahl-Hirschman Index computed based on the gross-output shares of macro-sectors; government consolidated gross debt as \% of GDP; government interest payable as \% of GDP; net government lending/borrowing as \% of GDP; gross fixed capital formation as \% of GDP; period average exchange rate; effective tax rate; and country ($c$) and year ($t$) fixed effects. All regressions in Panel A also include the ln of gross-output value ($pQ$).
\end{table}

\begin{table}[!h]
{\centering 
\caption{The replacement effect}
\label{tab:regr_replacement_lag}
\begin{adjustbox}{width=0.6\textwidth}\begin{tabular}{l ccc ccc}
\toprule \addlinespace
\csname @@input\endcsname{inputs/robustness/lagged_controls/Results3RICTint_cit_All_95t16}
\bottomrule
\end{tabular}\end{adjustbox}\\ }
\scriptsize
Notes: \sym{*}\(p<0.05\), \sym{**}\(p<0.01\), \sym{***}\(p<0.001\). Regression results to test the replacement effect for nineteen European countries during the period 1995-2016. All regressions use industry level data for the subset of industries susceptible to automation, defined as industries were the use of industrial robots is prevalent (see Appendix Table~\ref{appendix:data}). All regressions include: country industry ($ci$); country year ($ct$); and industry year ($it$) fixed effects. All regressions are weighted by the base-sample-year share of each industry’s number of hours worked to country-wide hours worked. Standard errors are clustered both at the country-industry and year level.
\end{table}

\begin{table}[!h]
{\centering 
\caption{The reinstatement effect}
\label{tab:regr_reinstatement_lagged_controls_Eurostat}
\begin{adjustbox}{width=0.9\textwidth}\begin{tabular}{l cccccc}
\toprule \addlinespace
\csname @@input\endcsname{inputs/robustness/lagged_controls/Results4shortRICTint_ct_All_95t16}
\bottomrule
\end{tabular}\end{adjustbox}\\ }
\scriptsize
Notes: $\ast$ \(p<0.05\), $\ast\ast$ \(p<0.01\), $\ast\ast\ast$ \(p<0.001\). Regression results to test the reinstatement effect for nineteen European countries during the period 2006-2016. All regressions include: GDP growth, government consolidated gross debt as \% of GDP; government interest payable as \% of GDP; net government lending/borrowing as \% of GDP; gross fixed capital formation as \% of GDP; period average exchange rate; value added TFP--calculated as the residual from an OLS regression of value-added volumes ($VA$) on a translog production function including capital volumes ($K$) and total number of hours worked ($L$); effective tax rate; and country ($c$) and year ($t$) fixed effects.
\end{table}

\begin{table}[!h]
{\centering 
\caption{The real income effect}
\label{tab:regr_realincome_lagged_controls_Eurostat}
\begin{adjustbox}{width=\textwidth}\begin{tabular}{l cccccccc}
\toprule \addlinespace
\csname @@input\endcsname{inputs/robustness/lagged_controls/Results5RICTint_ct_All_95t16}
\bottomrule
\end{tabular}\end{adjustbox}\\ \noindent }
\scriptsize
Notes: $\ast$ \(p<0.05\), $\ast\ast$ \(p<0.01\), $\ast\ast\ast$ \(p<0.001\). Regression results to test the real income effect for nineteen European countries during the period 2006-2016. Labor productivity is measured as the share of gross-output volumes ($Q$) over the total number of hours worked. TFP is calculated as the residual from an OLS regression of gross-output volumes ($Q$) on a translog production function including capital volumes ($K$), total number of hours worked ($L$) and intermediate input volumes ($M$). All regressions include: GDP growth, government consolidated gross debt as \% of GDP; government interest payable as \% of GDP; net government lending/borrowing as \% of GDP; gross fixed capital formation as \% of GDP; period average exchange rate; effective tax rate; and country ($c$) and year ($t$) fixed effects.
\end{table}

\newpage
\subsection{Using deeper lags ($t-1$, $t-2$ , $t-3$) of AT measures as IV}
\label{SM:lagged_iv}

\begin{table}[H]
\caption{Taxation and automation}
{\centering
\begin{adjustbox}{width=\textwidth}\begin{tabular}{l cccc|cccc|ccc}
\toprule \addlinespace
\csname @@input\endcsname{inputs/robustness/IVlags/Results2shortV1_RICTint_ct_All_95t16}
\bottomrule
\end{tabular}\end{adjustbox}} \\ 
\label{tab:regr_pre_tax_robots_IVlags} \scriptsize
Notes: $\ast$ \(p<0.05\), $\ast\ast$ \(p<0.01\), $\ast\ast\ast$ \(p<0.001\). Regression results of aggregate flows of tax revenues on different automation measures for nineteen European countries during the period 2003-2016. All regressions use country level data and include: GDP growth, gross output share of the service sector in total economy; Herfindahl-Hirschman Index computed based on the gross-output shares of macro-sectors; government consolidated gross debt as \% of GDP; government interest payable as \% of GDP; net government lending/borrowing as \% of GDP; gross fixed capital formation as \% of GDP; period average exchange rate; corporate tax rate; and country ($c$) and year ($t$) fixed effects. All regressions in Panel A also include the ln of gross-output value ($pQ$).
\end{table}

\begin{table}[H]
{\centering 
\caption{The replacement effect}
\label{tab:regr_replacement_ivlag}
\begin{adjustbox}{width=0.6\textwidth}\begin{tabular}{l ccc ccc}
\toprule \addlinespace
\csname @@input\endcsname{inputs/robustness/IVlags/Results3RICTint_cit_All_95t16}
\bottomrule
\end{tabular}\end{adjustbox}\\ }
\scriptsize
Notes: \sym{*}\(p<0.05\), \sym{**}\(p<0.01\), \sym{***}\(p<0.001\). Regression results to test the replacement effect for nineteen European countries during the period 1995-2016. All regressions use industry level data for the subset of industries susceptible to automation, defined as industries were the use of industrial robots is prevalent (see Appendix Table~\ref{appendix:data}). All regressions include: country industry ($ci$); country year ($ct$); and industry year ($it$) fixed effects. All regressions are weighted by the base-sample-year share of each industry’s number of hours worked to country-wide hours worked. Standard errors are clustered both at the country-industry and year level.
\end{table}

\begin{table}[H]
{\centering 
\caption{The reinstatement effect}
\label{tab:regr_reinstatement_IVlags}
\begin{adjustbox}{width=0.7\textwidth}\begin{tabular}{l cccccc}
\toprule \addlinespace
\csname @@input\endcsname{inputs/robustness/IVlags/Results4shortRICTint_ct_All_95t16}
\bottomrule
\end{tabular}\end{adjustbox}\\ }
\scriptsize
Notes: $\ast$ \(p<0.05\), $\ast\ast$ \(p<0.01\), $\ast\ast\ast$ \(p<0.001\). Regression results to test the reinstatement effect for nineteen European countries during the period 2003-2016. All regressions include: GDP growth, government consolidated gross debt as \% of GDP; government interest payable as \% of GDP; net government lending/borrowing as \% of GDP; gross fixed capital formation as \% of GDP; period average exchange rate; value added TFP--calculated as the residual from an OLS regression of value-added volumes ($VA$) on a translog production function including capital volumes ($K$) and total number of hours worked ($L$); corporate tax rate; and country ($c$) and year ($t$) fixed effects.
\end{table}

\begin{table}[H]
{\centering 
\caption{The real income effect}
\label{tab:regr_realincome_IVlags}
\begin{adjustbox}{width=0.9\textwidth}\begin{tabular}{l cccccccc}
\toprule \addlinespace
\csname @@input\endcsname{inputs/robustness/IVlags/Results5RICTint_ct_All_95t16}
\bottomrule
\end{tabular}\end{adjustbox}\\ \noindent }
\scriptsize
Notes: $\ast$ \(p<0.05\), $\ast\ast$ \(p<0.01\), $\ast\ast\ast$ \(p<0.001\). Regression results to test the real income effect for nineteen European countries during the period 2003-2016. Labor productivity is measured as the share of gross-output volumes ($Q$) over the total number of hours worked. TFP is calculated as the residual from an OLS regression of gross-output volumes ($Q$) on a translog production function including capital volumes ($K$), total number of hours worked ($L$) and intermediate input volumes ($M$). All regressions include: GDP growth, government consolidated gross debt as \% of GDP; government interest payable as \% of GDP; net government lending/borrowing as \% of GDP; gross fixed capital formation as \% of GDP; period average exchange rate; corporate tax rate; and country ($c$) and year ($t$) fixed effects.
\end{table}

\newpage
\subsection{Using trade data on imports of robot and ICT products to construct external IVs for the AT measures}
\label{SM:robustness_IV}

\begin{table}[H]
\caption{Taxation and automation}
{\centering
\begin{adjustbox}{width=0.9\textwidth}\begin{tabular}{l cccc|cccc|ccc}
\toprule \addlinespace
\csname @@input\endcsname{inputs/robustness/IV/Results2shortV1_RICTint_ct_All_95t16}
\bottomrule
\end{tabular}\end{adjustbox} \\ }
\label{tab:regr_pre_tax_robots_IV} \scriptsize
Notes: $\ast$ \(p<0.05\), $\ast\ast$ \(p<0.01\), $\ast\ast\ast$ \(p<0.001\). Regression results of aggregate flows of tax revenues on different automation measures for nineteen European countries during the period 1995-2016. All regressions use country level data and include: GDP growth, gross output share of the service sector in total economy; Herfindahl-Hirschman Index computed based on the gross-output shares of macro-sectors; government consolidated gross debt as \% of GDP; government interest payable as \% of GDP; net government lending/borrowing as \% of GDP; gross fixed capital formation as \% of GDP; period average exchange rate; imports as \% of GDP; exports as \% of GDP; and country ($c$) and year ($t$) fixed effects. All regressions in Panel A also include the ln of gross-output value ($pQ$).
\end{table}

\begin{table}[H]
{\centering 
\caption{The reinstatement effect}
\label{tab:regr_reinstatement_IV}
\begin{adjustbox}{width=0.7\textwidth,totalheight=0.80\textheight,keepaspectratio}\begin{tabular}{l cccccc}
\toprule \addlinespace
\csname @@input\endcsname{inputs/robustness/IV/Results4shortRICTint_ct_All_95t16}
\bottomrule
\end{tabular}\end{adjustbox}\\ }
\scriptsize
Notes: $\ast$ \(p<0.05\), $\ast\ast$ \(p<0.01\), $\ast\ast\ast$ \(p<0.001\). Regression results to test the reinstatement effect for nineteen European countries during the period 1995-2016. All regressions include: GDP growth, government consolidated gross debt as \% of GDP; government interest payable as \% of GDP; net government lending/borrowing as \% of GDP; gross fixed capital formation as \% of GDP; period average exchange rate; value added TFP--calculated as the residual from an OLS regression of value-added volumes ($VA$) on a translog production function including capital volumes ($K$) and total number of hours worked ($L$); imports as \% of GDP; exports as \% of GDP; and country ($c$) and year ($t$) fixed effects.
\end{table}

\begin{table}[H]
{\centering 
\caption{The real income effect}
\label{tab:regr_realincome_IV}
\begin{adjustbox}{width=0.90\textwidth}\begin{tabular}{l cccccccc}
\toprule \addlinespace
\csname @@input\endcsname{inputs/robustness/IV/Results5RICTint_ct_All_95t16}
\bottomrule
\end{tabular}\end{adjustbox}\\ \noindent }
\scriptsize
Notes: $\ast$ \(p<0.05\), $\ast\ast$ \(p<0.01\), $\ast\ast\ast$ \(p<0.001\). Regression results to test the real income effect for nineteen European countries during the period 1995-2016. Labor productivity is measured as the share of gross-output volumes ($Q$) over the total number of hours worked. TFP is calculated as the residual from an OLS regression of gross-output volumes ($Q$) on a translog production function including capital volumes ($K$), total number of hours worked ($L$) and intermediate input volumes ($M$). All regressions include: GDP growth, government consolidated gross debt as \% of GDP; government interest payable as \% of GDP; net government lending/borrowing as \% of GDP; gross fixed capital formation as \% of GDP; period average exchange rate; imports as \% of GDP; exports as \% of GDP; and country ($c$) and year ($t$) fixed effects.
\end{table}

\newpage
\section{Taxes, trade and income distribution}
\label{SM:robustness_taxes}

In this section, we provide the results from a series of robustness checks. First, we make sure that our results are not driven by changes in the tax system. Unfortunately, comprehensive data that covers the whole range of different taxes that is consistent across our sample of countries and covers a reasonable number of years is not available. 
We are only able to proxy tax reforms using data on corporate taxation that cover a smaller period of time but all countries in our sample. We use two different data sources. 

In \ref{SM:robustness_KPMG}, we repeat all baseline country level regressions and include as an additional control the corporate tax rate ($CRT_{c,t}$) sourced from KPMG.\footnote{The data were sourced from the KPMG website: \url{https://home.kpmg/xx/en/home/services/tax/tax-tools-and-resources/tax-rates-online.html}} 
This data are only available between 2003-2016 and only the results for the period after 2008 are comparable with the baseline analysis.
Next, in \ref{SM:robustness_ETR_Eurostat} we repeat all baseline country level regressions and include, as an additional control, the effective tax rate ($ERT_{c,t}$) sourced from Eurostat.\footnote{This is the Effective Average Tax Rate (ETR) for large corporations in non-financial sector, computed at corporate level, for average asset composition and funding sources, using the Devereux/Griffith 
methodology. The data are available in Eurostat: \url{https://ec.europa.eu/taxation_customs/business/economic-analysis-taxation/data-taxation_en}} 
The ETR variable is only available between 2006-2016 and again, only the results for the period after 2008 are comparable with the baseline analysis.

Another concern may arise from the impact of trade. 
To capture the country specific impact of trade, we repeat all baseline country level regressions and include, as additional controls, the country level imports ($Imports^{\% GDP}_{c,t}$) and exports ($Exports^{\% GDP}_{c,t}$) as percentage of GDP sourced from the OECD National Accounts Database.\footnote{Find data in OECD: \url{https://stats.oecd.org/viewhtml.aspx?datasetcode=NAAG\&lang=en\#}} 
Results are shown in Table \ref{SM:robustness_trade}. 

Furthermore, to explore the nexus between distribution and taxation, we examine the progressiveness of taxation (Table \ref{SM:robustness_progressive}). To do so, we rely on the same empirical specification used to understand the determinants of taxation, but now our regressions include, as an additional control, the Gini coefficient measuring cross-industry wage inequality ($Gini^{w}_{c,t}$) sourced from Eurostat.

Finally, we also test whether the attribution of capital taxes within the broad categories from OECD might affect our results. In particular we test whether the exclusion of the OECD tax category 1100 ``taxes on income, profits and capital gains of individuals'' from the taxes on capital $T^k$ changes the interpretation of our findings (assuming that these are neither capital nor labor taxes in the strict sense of their origin). The results are presented in \ref{SM:robustness_Tk} where we find similar results as in our baseline but the share of taxes as a percent of total taxes and GDP drops significantly due to this omission.

\subsection{Controlling for changes in corporate taxation using KPMG data}
\label{SM:robustness_KPMG}

\begin{table}[H]
{\centering
\caption{Taxation and automation}
\begin{adjustbox}{width=0.85\textwidth}
\begin{tabular}{l cccc|cccc|ccc}
\toprule \addlinespace
\csname @@input\endcsname{inputs/robustness/CTR_KPMG/Results2shortV1_RICTint_ct_All_95t16}
\bottomrule
\end{tabular}\end{adjustbox}\\}  
\label{tab:regr_pre_tax_robots_KPMG} \scriptsize
Notes: $\ast$ \(p<0.05\), $\ast\ast$ \(p<0.01\), $\ast\ast\ast$ \(p<0.001\). Regression results of aggregate flows of tax revenues on different automation measures for nineteen European countries during the period 2003-2016. All regressions use country level data and include: GDP growth, gross output share of the service sector in total economy; Herfindahl-Hirschman Index computed based on the gross-output shares of macro-sectors; government consolidated gross debt as \% of GDP; government interest payable as \% of GDP; net government lending/borrowing as \% of GDP; gross fixed capital formation as \% of GDP; period average exchange rate; corporate tax rate; and country ($c$) and year ($t$) fixed effects. All regressions in Panel A also include the ln of gross-output value ($pQ$).
\end{table}

\begin{table}[H]
{\centering 
\caption{The reinstatement effect}
\label{tab:regr_reinstatement_KPMG}
\begin{adjustbox}{width=0.7\textwidth}\begin{tabular}{l cccccc}
\toprule \addlinespace
\csname @@input\endcsname{inputs/robustness/CTR_KPMG/Results4shortRICTint_ct_All_95t16}
\bottomrule
\end{tabular}\end{adjustbox}\\ }
\scriptsize
Notes: $\ast$ \(p<0.05\), $\ast\ast$ \(p<0.01\), $\ast\ast\ast$ \(p<0.001\). Regression results to test the reinstatement effect for nineteen European countries during the period 2003-2016. All regressions include: GDP growth, government consolidated gross debt as \% of GDP; government interest payable as \% of GDP; net government lending/borrowing as \% of GDP; gross fixed capital formation as \% of GDP; period average exchange rate; value added TFP--calculated as the residual from an OLS regression of value-added volumes ($VA$) on a translog production function including capital volumes ($K$) and total number of hours worked ($L$); corporate tax rate; and country ($c$) and year ($t$) fixed effects.
\end{table}

\begin{table}[H]
{\centering 
\caption{The real income effect}
\label{tab:regr_realincome_KPMG}
\begin{adjustbox}{width=\textwidth}\begin{tabular}{l cccccccc}
\toprule \addlinespace
\csname @@input\endcsname{inputs/robustness/CTR_KPMG/Results5RICTint_ct_All_95t16}
\bottomrule
\end{tabular}\end{adjustbox}\\ \noindent }
\scriptsize
Notes: $\ast$ \(p<0.05\), $\ast\ast$ \(p<0.01\), $\ast\ast\ast$ \(p<0.001\). Regression results to test the real income effect for nineteen European countries during the period 2003-2016. Labor productivity is measured as the share of gross-output volumes ($Q$) over the total number of hours worked. TFP is calculated as the residual from an OLS regression of gross-output volumes ($Q$) on a translog production function including capital volumes ($K$), total number of hours worked ($L$) and intermediate input volumes ($M$). All regressions include: GDP growth, government consolidated gross debt as \% of GDP; government interest payable as \% of GDP; net government lending/borrowing as \% of GDP; gross fixed capital formation as \% of GDP; period average exchange rate; corporate tax rate; and country ($c$) and year ($t$) fixed effects.
\end{table}

\subsection{Controlling for changes in corporate taxation using Eurostat data}
\label{SM:robustness_ETR_Eurostat}

\begin{table}[H]
\caption{Taxation and automation}
{ \centering
\begin{adjustbox}{width=0.9\textwidth}
\centering
\begin{tabular}{l cccc|cccc|ccc}
\toprule \addlinespace
\csname @@input\endcsname{inputs/robustness/ETR_Eurostat/Results2shortV1_RICTint_ct_All_95t16}
\bottomrule
\end{tabular}\end{adjustbox} \\ }
\label{tab:regr_pre_tax_robots_ETR} \scriptsize
  Notes: $\ast$ \(p<0.05\), $\ast\ast$ \(p<0.01\), $\ast\ast\ast$ \(p<0.001\). Regression results of aggregate flows of tax revenues on different automation measures for nineteen European countries during the period 2006-2016. All regressions use country level data and include: GDP growth, gross output share of the service sector in total economy; Herfindahl-Hirschman Index computed based on the gross-output shares of macro-sectors; government consolidated gross debt as \% of GDP; government interest payable as \% of GDP; net government lending/borrowing as \% of GDP; gross fixed capital formation as \% of GDP; period average exchange rate; effective tax rate; and country ($c$) and year ($t$) fixed effects. All regressions in Panel A also include the ln of gross-output value ($pQ$).
\end{table}

\begin{table}[H]
{\centering 
\caption{The reinstatement effect}
\label{tab:regr_reinstatement_ETR_Eurostat}
\begin{adjustbox}{width=0.8\textwidth}\begin{tabular}{l cccccc}
\toprule \addlinespace
\csname @@input\endcsname{inputs/robustness/ETR_Eurostat/Results4shortRICTint_ct_All_95t16}
\bottomrule
\end{tabular}\end{adjustbox}\\ }
\scriptsize
Notes: $\ast$ \(p<0.05\), $\ast\ast$ \(p<0.01\), $\ast\ast\ast$ \(p<0.001\). Regression results to test the reinstatement effect for nineteen European countries during the period 2006-2016. All regressions include: GDP growth, government consolidated gross debt as \% of GDP; government interest payable as \% of GDP; net government lending/borrowing as \% of GDP; gross fixed capital formation as \% of GDP; period average exchange rate; value added TFP--calculated as the residual from an OLS regression of value-added volumes ($VA$) on a translog production function including capital volumes ($K$) and total number of hours worked ($L$); effective tax rate; and country ($c$) and year ($t$) fixed effects.
\end{table}

\begin{table}[H]
{\centering 
\caption{The real income effect}
\label{tab:regr_realincome_ETR_Eurostat}
\begin{adjustbox}{width=\textwidth}\begin{tabular}{l cccccccc}
\toprule \addlinespace
\csname @@input\endcsname{inputs/robustness/ETR_Eurostat/Results5RICTint_ct_All_95t16}
\bottomrule
\end{tabular}\end{adjustbox}\\ \noindent }
\scriptsize
Notes: $\ast$ \(p<0.05\), $\ast\ast$ \(p<0.01\), $\ast\ast\ast$ \(p<0.001\). Regression results to test the real income effect for nineteen European countries during the period 2006-2016. Labor productivity is measured as the share of gross-output volumes ($Q$) over the total number of hours worked. TFP is calculated as the residual from an OLS regression of gross-output volumes ($Q$) on a translog production function including capital volumes ($K$), total number of hours worked ($L$) and intermediate input volumes ($M$). All regressions include: GDP growth, government consolidated gross debt as \% of GDP; government interest payable as \% of GDP; net government lending/borrowing as \% of GDP; gross fixed capital formation as \% of GDP; period average exchange rate; effective tax rate; and country ($c$) and year ($t$) fixed effects.
\end{table}

\subsection{Controlling for trade}
\label{SM:robustness_trade}

\begin{table}[H]
\caption{Taxation and automation}
{ \centering
\begin{adjustbox}{width=0.85\textwidth}
\centering
\begin{tabular}{l cccc|cccc|ccc}
\toprule \addlinespace
\csname @@input\endcsname{inputs/robustness/Trade_OECD/Results2shortV1_RICTint_ct_All_95t16}
\bottomrule
\end{tabular}\end{adjustbox} \\ }
\label{tab:regr_pre_tax_robots_Trade} \scriptsize
Notes: $\ast$ \(p<0.05\), $\ast\ast$ \(p<0.01\), $\ast\ast\ast$ \(p<0.001\). Regression results of aggregate flows of tax revenues on different automation measures for nineteen European countries during the period 1995-2016. All regressions use country level data and include: GDP growth, gross output share of the service sector in total economy; Herfindahl-Hirschman Index computed based on the gross-output shares of macro-sectors; government consolidated gross debt as \% of GDP; government interest payable as \% of GDP; net government lending/borrowing as \% of GDP; gross fixed capital formation as \% of GDP; period average exchange rate; imports as \% of GDP; exports as \% of GDP; and country ($c$) and year ($t$) fixed effects. All regressions in Panel A also include the ln of gross-output value ($pQ$).
\end{table}

\begin{table}[H]
{\centering 
\caption{The reinstatement effect}
\label{tab:regr_reinstatement_trade}
\begin{adjustbox}{width=0.8\textwidth}\begin{tabular}{l cccccc}
\toprule \addlinespace
\csname @@input\endcsname{inputs/robustness/Trade_OECD/Results4shortRICTint_ct_All_95t16}
\bottomrule
\end{tabular}\end{adjustbox}\\ }
\scriptsize
Notes: $\ast$ \(p<0.05\), $\ast\ast$ \(p<0.01\), $\ast\ast\ast$ \(p<0.001\). Regression results to test the reinstatement effect for nineteen European countries during the period 1995-2016. All regressions include: GDP growth, government consolidated gross debt as \% of GDP; government interest payable as \% of GDP; net government lending/borrowing as \% of GDP; gross fixed capital formation as \% of GDP; period average exchange rate; value added TFP--calculated as the residual from an OLS regression of value-added volumes ($VA$) on a translog production function including capital volumes ($K$) and total number of hours worked ($L$); imports as \% of GDP; exports as \% of GDP; and country ($c$) and year ($t$) fixed effects.
\end{table}

\begin{table}[!h]
{\centering 
\caption{The real income effect}
\label{tab:regr_realincome_trade}
\begin{adjustbox}{width=\textwidth}\begin{tabular}{l cccccccc}
\toprule \addlinespace
\csname @@input\endcsname{inputs/robustness/Trade_OECD/Results5RICTint_ct_All_95t16}
\bottomrule
\end{tabular}\end{adjustbox}\\ \noindent }
\scriptsize
Notes: $\ast$ \(p<0.05\), $\ast\ast$ \(p<0.01\), $\ast\ast\ast$ \(p<0.001\). Regression results to test the real income effect for nineteen European countries during the period 1995-2016. Labor productivity is measured as the share of gross-output volumes ($Q$) over the total number of hours worked. TFP is calculated as the residual from an OLS regression of gross-output volumes ($Q$) on a translog production function including capital volumes ($K$), total number of hours worked ($L$) and intermediate input volumes ($M$). All regressions include: GDP growth, government consolidated gross debt as \% of GDP; government interest payable as \% of GDP; net government lending/borrowing as \% of GDP; gross fixed capital formation as \% of GDP; period average exchange rate; imports as \% of GDP; exports as \% of GDP; and country ($c$) and year ($t$) fixed effects.
\end{table}

\subsection{The progressiveness of taxation}
\label{SM:robustness_progressive}

\begin{table}[H]
{\centering 
\caption{Taxation and the structure of economic production}
\begin{adjustbox}{width=\textwidth}
\begin{tabular}{l cccc|cccc|ccc}
\toprule \addlinespace
\csname @@input\endcsname{inputs/robustness/TaxProgressiveness/Results2RICTint_ct_All_95t16}
\bottomrule
\end{tabular}\end{adjustbox} \\ }
\scriptsize
Notes: $\ast$ \(p<0.05\), $\ast\ast$ \(p<0.01\), $\ast\ast\ast$ \(p<0.001\). Regression results to establish the link between tax aggregation and economic production for nineteen European countries during the period 1995-2016. All regressions use country level data and include: GDP growth ($GDPgrowth_{c,t}$), share of gross output produced in service industries ($Services_{c,t}$); the Herfindahl-Hirschman Index computed based on the gross-output shares of macro-sectors ($HHI_{c,t}$); government consolidated gross debt as \% of GDP ($Debt^{\%GDP}_{c,t}$); government interest payable as \% of GDP ($Interest^{\%GDP}_{c,t}$); net government lending/borrowing as \% of GDP ($Lending^{\%GDP}_{c,t}$); gross fixed capital formation as \% of GDP ($GovInv^{\%GDP}_{c,t}$); period average exchange rate ($XRate_{c,t}$); Gini index from the industry level distribution of hourly wage ($Gini^{w}_{c,t}$); and country ($c$) and year ($t$) fixed effects. For the first block, $wL_{c,t}$, $rK_{c,t}$ and $pQ_{c,t}$ are expressed as the natural logarithm ($ln$) while for the last two blocks they are expressed as \% of GDP.
\end{table}

\subsection{Excluding OECD group 1100 ``taxes on income, profits and capital gains of individuals'' from capital taxes}
\label{SM:robustness_Tk}

\begin{table}[H]
\caption{Taxation and automation}
{\centering
\begin{adjustbox}{width=0.85\textwidth}
\centering
\begin{tabular}{l cccc|cccc|ccc}
\toprule \addlinespace
\csname @@input\endcsname{inputs/robustness/Tk_alt/Results2shortV1_RICTint_ct_All_95t16}
\midrule
\csname @@input\endcsname{inputs/robustness/Tk_alt/Results2shortV1_RICTint_ct_All_95t07}
\midrule
\csname @@input\endcsname{inputs/robustness/Tk_alt/Results2shortV1_RICTint_ct_All_08t16}
\bottomrule
\end{tabular}\end{adjustbox} \\ }
\label{tab:regr_pre_tax_robots_Trade} \scriptsize
Notes: $\ast$ \(p<0.05\), $\ast\ast$ \(p<0.01\), $\ast\ast\ast$ \(p<0.001\). Regression results of aggregate flows of tax revenues on different automation measures for nineteen European countries during the period 1995-2016. All regressions use country level data and include: GDP growth, gross output share of the service sector in total economy; Herfindahl-Hirschman Index computed based on the gross-output shares of macro-sectors; government consolidated gross debt as \% of GDP; government interest payable as \% of GDP; net government lending/borrowing as \% of GDP; gross fixed capital formation as \% of GDP; period average exchange rate; and country ($c$) and year ($t$) fixed effects. All regressions in Panel A also include the ln of gross-output value ($pQ$). Taxes on capital $T^k$ exclude the OECD tax category 1100 ``taxes on income, profits and capital gains of individuals''.
\end{table}

\newpage
\begin{spacing}{0.8}
\putbib 
\end{spacing}
\end{bibunit}

\end{document}